\documentclass[prd,fleqn,eqsecnum,nofootinbib,preprint,
               tightenlines]{revtex4}

\usepackage{amsmath}
\usepackage{amsfonts}
\usepackage{latexsym}
\usepackage{mathrsfs}
\usepackage{graphicx}

\newcommand{\bR}{\mathbb{R}}
\newcommand{\bC}{\mathbb{C}}
\newcommand{\bZ}{\mathbb{Z}}

\def\D{\mathcal{D}}

\def\d{{\rm d}}
\newcommand{\half}{{\textstyle \frac{1}{2}}}

\renewcommand{\i}{{\rm i}}

\newcommand{\fdq}[2]{\frac{\delta #1}{\delta #2}}

\newcommand{\dx}{\!\!{\rm d}^4x\,\,}

\newcommand{\tr}{\text{tr}}
\newcommand{\eq}{=}
\newcommand{\twovect} [1]{\left( \begin{array}{c} #1 \end{array} \right) }
\newcommand{\vect}    [1]{\left( \begin{array}{c} #1 \end{array} \right)}
\newcommand{\vectleft}[1]{\left( \begin{array}{l} #1 \end{array} \right)}
\newcommand{\twomat}  [1]{\left( \begin{array}{cc} #1 \end{array} \right)}

\newcommand{\Gcl}{\Gamma_{\rm cl}}
\newcommand{\id}{\openone}
\newcommand{\Dsl}{D\!\!\!\!/\,}

\newcommand{\bcs}  {boundary conditions}

\newcommand{\YMH}  {Yang--Mills--Higgs}

\newcommand{\SM}   {Standard Model}

\newcommand{\ewSM} {electroweak Standard Model}
\begin{document}

\noindent J. Math. Phys., Vol. 44, No. 8, August 2003
\hfill KA--TP--01--2003 \newline
\mbox{}\hfill hep-th/0304167\newline
\vspace{1\baselineskip}
\title{\Large \bf Sphalerons, spectral flow, and anomalies}

\author{Frans R.\ Klinkhamer}
\email{frans.klinkhamer@physik.uni-karlsruhe.de}
\author{Christian Rupp}
\email{cr@particle.uni-karlsruhe.de}
\affiliation{Institut f\"ur Theoretische Physik,
             Universit\"at Karlsruhe (TH),\\76128 Karlsruhe, Germany\\ \\}

\begin{abstract}
\vspace{0.5\baselineskip}\noindent
The topology of configuration space may be responsible
in part for the existence of sphalerons. Here, sphalerons are defined to be
static but unstable finite-energy solutions of the classical field equations.
Another manifestation of the nontrivial topology of configuration
space is the phenomenon of spectral flow for the eigenvalues of the
Dirac Hamiltonian. The spectral flow, in turn, is related to the
possible existence of anomalies. In this review,
the interconnection of these topics is illustrated for three
particular sphalerons of $SU(2)$ Yang--Mills--Higgs theory.
\vspace{2\baselineskip}\newline
Invited paper for the special issue of the Journal of Mathematical Physics
on \emph{``Integrability, topological solitons and beyond''}
edited by T. Fokas and N.S. Manton.
\end{abstract}


\maketitle

\section{Introduction}
\label{sec:intro}

One of the main themes of the present special issue
concerns the so-called topological solitons. The field configurations
of these classical solutions are characterized by a
topologically non\-trivial map of the space manifold (or part of it)
into some in\-ter\-nal space of the model considered. A well-known example
is the Skyrme soliton \cite{Sk60}, for which the space manifold
$S^3$ (i.e., the compactified Euclidean space $\bR^3$) is mapped
into the internal space $SU(2)$. Another example is the magnetic
monopole \cite{tHP74}, for which the ``sphere at infinity'' $S^2$ is
mapped into the Higgs vacuum manifold $SO(3)/SO(2)$.

There exist, however, other classical solutions, the so-called
sphalerons, which themselves have trivial topology but trace back
to nontrivial topology in the configuration space of the fields
\cite{T82,M83}. In this contribution, we intend to give an elementary
discussion of sphaleron solutions in Yang--Mills--Higgs theory and
the underlying topology. In order to get a clear picture of what goes on,
we focus on a single Yang--Mills--Higgs theory and three specific
sphalerons \cite{KM84,K93,KO94}.

Physically, the topological solitons and the sphalerons play a
different role.  Solitons are primarily relevant to the
equilibrium properties of the theory (e.g., the existence of certain
stable asymptotic states), whereas sphalerons are of importance to the
dynamics of the theory.
The sphaleron \cite{KM84} of the electroweak Standard Model \cite{EWSM},
for example, is believed to play a crucial role for baryon-number-violating
processes in the early universe (see, e.g., Refs.\
\cite{McL94,RS96} for two reviews).

The outline of this article is as follows. In Section \ref{sec:YMH}, we present
the theory considered, to wit, $SU(2)$ Yang--Mills theory with a single
complex isodoublet of Higgs fields. This particular Yang--Mills--Higgs
theory forms the core of the electroweak Standard Model of elementary
particle physics.
In Section \ref{sec:spheres}, we recall some basic facts about the mapping of
spheres into spheres, in particular their homotopy classes.
In Section \ref{sec:sphalerons}, we describe three sphaleron solutions and
their topological \emph{raison d'\^etre}.
In Section \ref{sec:sflow}, we discuss another manifestation of the nontrivial
topology of configuration space, namely the spectral flow of the eigenvalues of
the Dirac Hamiltonian. The word ``spectral flow'' is used in a generalized sense,
meaning \emph{any} type of rearrangement of the energy levels.
Loosely speaking, the spectral flow makes it possible for a
sphaleron to acquire a fermion zero-mode.
In Section \ref{sec:anomalies}, we link
the spectral flow to the possible occurrence of anomalies
(which signal the loss of one or more classical symmetries).
In Section \ref{sec:conclusion}, finally, we present
some concluding remarks.

\section{$SU(2)$ Yang--Mills--Higgs theory}
\label{sec:YMH}
In this article, we consider a simplified version of the
electroweak Standard Model \cite{EWSM} without the hypercharge $U(1)$
gauge field.  This means, basically, that we set the weak mixing
angle $\theta_w \equiv \arctan(g'/g)$ to zero, where $g'$ and $g$
are the coupling constants of the $U(1)$ and $SU(2)$ gauge groups,
respectively. Also, we take only one family of quarks and
leptons instead of the three known experimentally.

In general, the fields are considered to propagate in Minkowski spacetime
with coordinates $x^\mu \in \bR$, $\mu=0,1,2,3$, and metric
$g_{\mu\nu}(x)=\mathrm{diag}(+1,-1,-1,-1)$.
But occasionally we go over to Euclidean spacetime with metric
$g_{\mu\nu}(x)= \delta_{\mu\nu}$.
Natural units with $\hbar$ $=$ $c$ $=$ $1$ are used throughout.

The $SU(2)$ Yang--Mills gauge field is denoted by
$A_\mu (x) \equiv A_\mu^a (x)\, \tau^a / (2\i)$, where the $\tau^a$ are the
three Pauli matrices acting on weak isospin space
and the component fields $A_\mu^a (x)$ are real.
(Repeated indices are summed over, unless stated otherwise.)
The complex Higgs field transforms as an isodoublet
under the $SU(2)$ gauge group and is given by $\Phi (x)=
\left(\Phi_1 (x), \Phi_2 (x) \right)^{\,t}$,
where the suffix $t$ stands for transpose [cf.
Eq. (\ref{PhiMdef}) below]. The fermion fields will be discussed in Section
\ref{sec:sflow}.

The classical action of the gauge and Higgs fields reads
\begin{equation}
\Gamma_{\rm YMH} = \int_{\bR^4}\; \dx
\left\{ \half \, \tr\, F_{\mu\nu}  F^{\mu\nu} +
       (D_\mu  \Phi)^\dagger (D^\mu \Phi)     -
       \lambda \, \Bigl( \Phi^\dagger \Phi - \eta^2 \Bigr)^2 \right\}\,,
\label{actionYMH}
\end{equation}
where $F_{\mu\nu}\equiv\partial_\mu A_\nu -\partial_\nu A_\mu + g [A_\mu,
A_\nu]$ is the  $SU(2)$ Yang--Mills field strength and
$D_\mu \Phi \equiv (\partial_\mu + g A_\mu) \Phi$
the covariant derivative of the Higgs field. The theory
has Yang--Mills coupling constant $g$ and quartic Higgs coupling constant
$\lambda$, but the classical dynamics depends only on the ratio
$\lambda/g^2$. The parameter $\eta$ has the dimension of mass and
sets the scale of the Higgs expectation value.
The three $W$ vector bosons then have equal mass,
$M_W= g\, \eta / \sqrt{2}$.
The single Higgs scalar boson has a mass $M_H=2\,\sqrt{\lambda}\:\eta \,$.

The action (\ref{actionYMH}) is invariant under a local gauge transformation
\begin{equation}
\Phi^\prime (x)    = \Lambda(x)\, \Phi(x)\,, \qquad
g  A^\prime_\mu(x) = \Lambda(x)\, \bigl( \,g A_\mu(x)+
                     \partial_\mu  \, \bigr)\,\Lambda(x)^{-1}\,,
\end{equation}
for an arbitrary gauge function $\Lambda(x) \in SU(2)$.
In addition, there are certain global $SU(2)$ and $U(1)$ symmetry
transformations which operate solely on the Higgs field.

\section{Maps of spheres into spheres}\label{sec:spheres}

Let us consider continuous maps from a
connected manifold $M$ to a connected manifold $N$.
Two such maps, $f_1$ and $f_2$,  are called \emph{homotopic}
if the one can be obtained from the other by continuous deformation.
More specifically, $f_1$ and $f_2$ are homotopic if there exists a
continuous map $h: [0,1] \times M \to N$ such that $h(0, m)=f_1(m)$
and $h(1,m)=f_2(m)$ for all $m \in M$.
All maps $M\to N$ can be divided into equivalence classes, where two
maps are equivalent if they are homotopic (see, e.g., Ref.~\cite{Nak90}).

\begin{figure}[t]
\begin{center}
\includegraphics[height=3cm]{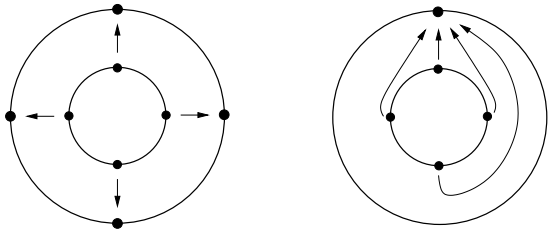}
\end{center}
\caption{Two nonhomotopic maps $S^1\to S^1$, with inner circles
  mapped into outer circles and matching points indicated. For the
  figure on the right, the whole inner circle is mapped into a single
  point of the outer circle.}
\label{fig:circ}
\end{figure}

We are particularly interested in the case where $M$ and $N$ are the spheres
$S^m$ and $S^n$, respectively. The set of
homotopy classes is called the \emph{homotopy group} and is denoted by
$\pi_m(S^n)$. Figure~\ref{fig:circ} shows two maps $S^1 \to S^1$
which are not homotopic. It is clear that in this particular case the
homotopy classes can be labeled by integer numbers which describe how
often the original circle $S^1$ is wrapped around the target circle $S^1$.
This explains the result $\pi_1(S^1)=\bZ$, where $\bZ$ denotes the group
of integers under addition.
The two maps shown in Fig.\ \ref{fig:circ} have winding numbers $1$ and $0$.

The homotopy classes of $S^n\to S^n$, for $n>1$, can be pictured analogously,
since the representation of a sphere $S^n$ in spherical coordinates contains
exactly one azimuthal angle $\phi \in [0,2\pi]$. The result is $\pi_n(S^n) = \bZ$.
Further homotopy groups are:
$\pi_m(S^n)=\{0\}$ for $m<n$, $\pi_3(S^2)=\bZ$, and $\pi_4(S^3)=\bZ_2$,
where $\bZ_2$ denotes the group
of integers $\{0,1\}$ under addition modulo 2.

Next, consider families of maps $S^m \to S^n$, where the family parameters themselves
form a sphere $S^p$. In short, consider $S^p \times S^m \to S^n$.
Imposing certain constraints, these families of maps
can be viewed as maps $S^{p+m}\to S^n$ and classified according to the
homotopy groups of spheres.

To this end, we introduce the \emph{smash product} \cite{Nak90}
of two spheres $S^p$ and $S^m$.
The smash product $S^p \wedge S^m$ is defined as
the Cartesian product $S^p \times S^m$ with
the set $(\{x_0\} \times S^m) \cup (S^p \times \{y_0\})$ considered as a
single point, for some arbitrarily chosen $x_0 \in S^p$ and $y_0 \in S^m$.
It can be shown that $S^p \wedge S^m$ is homeomorphic to the sphere $S^{p+m}$
(see Fig.~\ref{fig:smash} for a sketch of the proof).

A simple corollary will be important in the following.
Any map $f: S^p \times S^m \to S^n$ can effectively be considered as a map
defined on $S^p \wedge S^m$ if $f(x_0, y)$ is independent of $y$ and
$f(x, y_0)$ is independent of $x$, for an appropriate choice of
$x_0 \in S^p$ and $y_0 \in S^m$.

\newcommand{\figcap}{Left: Cartesian product space $S^1 \times S^1$
drawn as a torus, with two circles representing the sets
$\{x_0\}\times S^1$ and $S^1 \times \{y_0\}$ for some arbitrarily
chosen points $x_0$ and $y_0$.
Middle: surface after shrinking the first of these sets to a point.
Right:  surface after shrinking the second set to a point, which
gives the smash product $S^1\wedge S^1 \sim S^2$.}
\begin{figure*}[t]
\begin{center}
\includegraphics{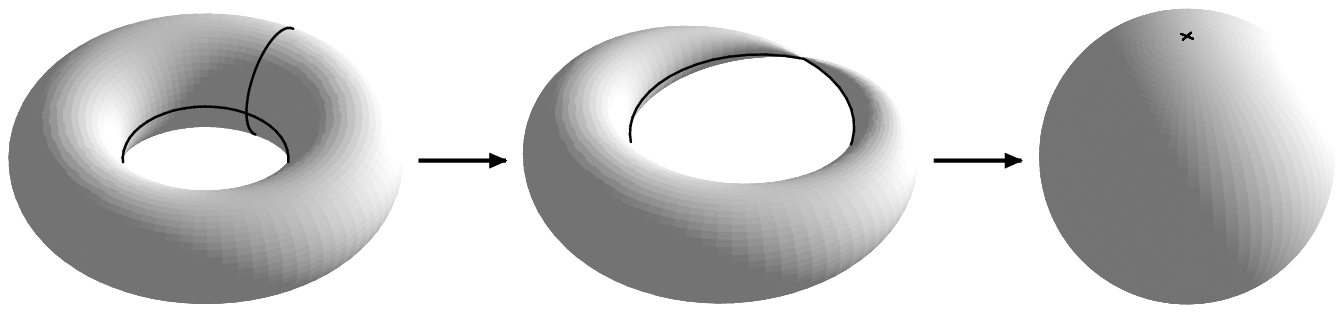}
\end{center}
\caption{\label{fig:smash}\protect\figcap}
\end{figure*}

\section{Sphalerons}
\label{sec:sphalerons}
The word sphaleron is of Greek origin and means ``ready to fall''
(see Ref.~\cite{KM84}~for the et\-y\-mol\-o\-gy).
It is used to denote a static but unstable solution of
the classical field equations with finite total energy of the fields.

In this article, only finite-energy configurations of the fields
will be considered. By analogy to
Morse theory \cite{Mil63}, sphalerons can then be looked for by a minimax
procedure  \cite{T82} if the configuration space of the underlying field theory
is multiply connected.

The procedure runs as follows: first, construct a noncontractible
$p$-dimensional sphere $S^p$ in configuration space, then determine its
maximal energy configuration, and, finally, ``shrink'' the sphere
to minimize this maximal energy. If the configuration space were compact,
this procedure  would be guaranteed to give a saddle
point. But configuration space is infinite-dimensional and noncompact,
so that the minimax procedure produces at best only a candidate solution.
It has to be checked explicitly that the appropriate minimax-configuration
solves the classical field equations. If this is the case, the minimax-configuration
is a genuine sphaleron.

\subsection{Sphaleron $\mathrm{S}$}
\label{sec:sphaleronS}

For the sphaleron $\mathrm{S}\,$ \cite{M83,KM84,DHN74}
of the $SU(2)$ Yang--Mills--Higgs theory (\ref{actionYMH}),
we consider three-space to be compactified by adding the ``sphere at
infinity.'' Configuration space is then the space of all
static three-dimensional gauge and Higgs field configurations in a
particular gauge which have finite energy.
The static gauge field can be written as a Lie-algebra-valued one-form,
\begin{equation}
A(x^1,x^2,x^3) \equiv A_m^a (x^1,x^2,x^3)\,\tau^a / (2\i)\, {\rm d}x^m \,,
\end{equation}
with implicit sums of $a$ and $m$ over $1,\,2,\,3$.
Furthermore, we use spherical coordinates $(r,\theta,\phi)$ over $\bR^3$
and employ the radial gauge condition $A_r=0$, together with $A_0 =0$.

Since the energy has to
be finite, only those configurations are admissible for which the gauge
field tends towards a pure-gauge configuration as $r \to \infty$
and the Higgs field towards its associated vacuum value,
\begin{equation}
g\, A^\infty = -  \d \omega\, \omega^{-1}\,, \qquad
\Phi^\infty = \eta \: \omega \twovect{0\\1}\,,
\end{equation}
for  a map $\omega$ of the ``sphere at infinity'' $S^2$ into the gauge
group $SU(2)$.

Any loop in configuration space
induces a loop in the space of these mappings $\omega$. The corresponding map
is denoted by
\begin{equation} \label{UNCL}
U: S^1 \times S^2 \to SU(2)\,, \qquad (\mu, \theta, \phi) \mapsto U(\mu,
\theta, \phi)\,,
\end{equation}
where $\mu \in [0,\pi]$ is the parameter of the loop of configurations
and $\theta\in [0,\pi]$ and  $\phi\in [0,2\pi]$ are
spherical coordinates in three-space.

By imposing certain constraints on this
map, we may effectively reduce the set of allowed loops,
so that $U$ becomes a map $S^3 \to SU(2) \sim S^3$ which falls into homotopy
classes according to $\pi_3(SU(2))=\pi_3(S^3)=\bZ$.
To be specific, the map $U$ for $\mu\eq 0$ and $\mu \eq \pi$ must not depend
on $\theta$ and $\phi$, and the map  $U$ for $\theta\eq 0$ has to be independent
of $\mu$. Then $U$ is effectively defined on the smash product
$S^1 \wedge S^2 \sim S^3$,
as explained in the last paragraph of Section \ref{sec:spheres}.
Now there exist noncontractible loops of field configurations
for which the minimax procedure can be performed.

An appropriate expression for the map (\ref{UNCL}) is given by \cite{M83}
\begin{equation} \label{U}
U(\mu, \theta, \phi) \eq  y^1 \,(-\i \tau_1) + y^2 \,(-\i \tau_2) +
                          y^3 \,(-\i \tau_3) + y^4 \,\id_2 \,,
\end{equation}
with
\begin{equation} \label{y}
\vect{y^1\\y^2\\y^3\\y^4} \eq
\vect{
-\sin\mu \sin\theta \sin\phi\\
-\sin\mu \sin\theta \cos\phi\\
\sin\mu\cos\mu \,(\cos\theta-1)\\
\cos^2\mu + \sin^2 \mu \cos\theta
} \,.
\end{equation}
In order to calculate the winding number of this particular map $U$,
we examine its relation to the standard spherical coordinates on $S^3$,
\begin{equation}
\vect{z^1\\z^2\\z^3\\z^4} \eq
 \vectleft{
\cos\theta_2\\
\sin\theta_2\cos\theta_1\\
\sin\theta_2 \sin\theta_1\sin\phi_1\\
\sin\theta_2 \sin\theta_1\cos\phi_1
}\,,
\end{equation}
with polar angles $\theta_1,\theta_2 \in [0,\pi]$ and
azimuthal angle $\phi_1\in [0,2\pi]$.

We first observe that the two-vector
\begin{equation} \vec w \eq \vect{\cos\mu\\ \sin\mu \cos\theta} \end{equation}
sweeps over the unit disk if $\mu$ and  $\theta$ run from $0$ to $\pi$.
Since rotations map the unit disk onto itself and leave the length of
$\vec w$ invariant,
\begin{equation}
|\vec w|^2 \eq 1-\sin^2\mu \,\sin^2\theta \,,
\end{equation}
the relation
\begin{equation}
\vect{\cos\theta_2 \\ \sin\theta_2 \cos\theta_1} \eq
\twomat{\cos\alpha  & - \sin\alpha \\ \sin\alpha & \cos\alpha}
\vect{\cos\mu \\ \sin\mu \cos\theta}
\end{equation}
describes an admissible reparametrization of the disc.
By choosing $\alpha \eq -\mu$, we find $y^4 \eq z^1$ and $y^3 \eq z^2$.
With $\sin\theta_1 \sin\theta_2 \eq \sin\mu\sin\theta$ and $\phi_1 \eq \phi$,
we have also $y^1\eq -z^3$ and $y^2\eq -z^4$.

The conclusion is that the
map $U$ as defined by Eqs.~(\ref{U})--(\ref{y}) covers the target sphere $S^3$
exactly once.  The map $U$ has winding number one
(or minus one, depending on the definition of the winding number)
and corresponds to a  nontrivial element of the homotopy group
$\pi_3(S^3)=\bZ$.

For the static $SU(2)$ gauge and Higgs fields of the noncontractible loop
(NCL), we make the \emph{Ansatz} \cite{M83}
\begin{subequations}
\label{NCL_S}
\mathindent=0cm
\begin{align}
g\, A(\mu, r, \theta, \phi) &= -  \,f(r)\, \,
    \d U(\mu, \theta, \phi) \,\, U(\mu, \theta, \phi)^{-1}\,,\\
\Phi(\mu, r, \theta, \phi) &=
    \eta \, h(r) \,U(\mu, \theta, \phi) \vect{0\\1} +
    \eta \, \Bigl( 1-h(r) \Bigr) \vect{0\\ \exp(-\i\,\mu) \cos\mu} ,
\end{align}
\end{subequations}
with the following boundary conditions for the
radial functions $f$ and $h$:
\begin{subequations} \label{NCL_bcs}
\begin{align}
\lim_{r\to 0} \, f(r)/r &=0\,, & \lim_{r\to\infty} \, f(r)&=1\,, \\
\lim_{r\to 0} \, h(r)&=0\,, & \lim_{r\to\infty} \, h(r)&=1\,.
\end{align}
\end{subequations}
The energy density of the fields (\ref{NCL_S})
turns out to be spherically symmetric.
Indeed, the fields of the NCL can also be written in a manifestly
spherically symmetric form \cite{K90plb}.

\begin{figure}[t]
\begin{center}
\includegraphics[width=5cm]{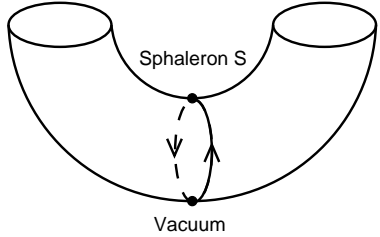}
\end{center}
\caption{Sphaleron $\mathrm{S}$ on top of a noncontractible loop in
  configuration space. The field energy is zero for the vacuum and
  positive for the sphaleron $\mathrm{S}$.}
\label{fig:sphaleronS}
\end{figure}

The fields (\ref{NCL_S}) of the NCL at $\mu=0$ or $\pi$ correspond to
the Higgs vacuum with $A(x)=0$ and $\Phi(x)= (0,\eta)^{\,t}$.
This particular configuration is independent of the radial functions
$f$ and $h$ and has zero energy according to Eq.~(\ref{actionYMH}).
The NCL configuration at $\mu= \pi/2$
is distinguished by having parity reflection symmetry
(the only other configuration  of the NCL with this property is
the vacuum at $\mu=0$). For given functions $f$ and $h$,
this $\mu= \pi/2$ configuration is also the
maximum energy configuration over the NCL.
The minimax procedure now consists of adjusting the radial
functions $f$ and $h$ while maintaining the boundary conditions (\ref{NCL_bcs}),
so that the energy at $\mu=\pi/2$ is
minimized. The resulting configuration is the sphaleron
$\mathrm{S}$, as sketched in Fig.~\ref{fig:sphaleronS}

Using numerical methods, one finds for the sphaleron energy the value
\cite{KM84}
\begin{equation} \label{ES}
E_\mathrm{S} \approx 3.04 \times (4\pi /g^2)\, M_W  \,,
\end{equation}
which holds for the case of vanishing quartic Higgs coupling
constant ($\lambda/g^2 =0$). [The sphaleron energy
$E_\mathrm{S}$ has also been calculated for the full
$SU(2) \times U(1)$ Yang--Mills--Higgs theory of the electroweak Standard
Model. The energy $E_\mathrm{S}$ is found to be weakly dependent
on the mixing angle $\theta_w$, at least near $\theta_w =0$.
The emergence of a large magnetic dipole moment
$\mu_\mathrm{S} \propto (4\pi/g^2)\, g\, \tan\theta_w /M_W$ is perhaps
more interesting. See Refs.\ \cite{KM84,KKB92,HJ94} for details.]
For large enough values of $\lambda/g^2$, additional solutions appear, the
so-called ``deformed sphalerons'' \cite{KB89,Y89}. The appearance of these
extra sphalerons can be explained  \cite{K90plb} by a simple deformation
of the energy surface in Fig.~\ref{fig:sphaleronS}.

The sphaleron $\mathrm{S}\,$  by itself has trivial topology, with the
``sphere at infinity'' $S^2$ mapped into the Higgs vacuum manifold
$SU(2)\sim S^3$; cf. Section \ref{sec:spheres}.
(As mentioned in the Introduction, the magnetic monopole \cite{tHP74}
in $SU(2)$ Yang--Mills theory with a real isotriplet of Higgs
is based on the nontrivial map $S^2 \to SO(3)/SO(2) \sim S^2$.)
Note that the original $\mathrm{S}\,$ \emph{Ansatz},
with the so-called hedgehog structure, was discovered \cite{DHN74} ten
years before the construction of $\mathrm{S}\,$ via the NCL \cite{M83,KM84}.

In the radial gauge, the vacuum configuration of the $SU(2)$ gauge field
is uniquely fixed, $g A_m^{\rm vac}(x^1,x^2,x^3)=0$. If this gauge condition
is abandoned, any pure-gauge configuration $g A_m= - \partial_m U \,
U^{-1}$, for arbitrary time-independent $SU(2)$-valued field $U$, is a
possible vacuum configuration. Depending on the topology of three-space,
these vacuum configurations may or may not fall into different unconnected
classes. This does not happen for our compactification $\bR^3
\cup S^2_\infty$. But the situation changes
if, instead, we choose a one-point compactification
$\bR^3\cup \{\infty\}$, with all fields approaching a single
direction-independent value as $r\to \infty$.
Each vacuum configuration then corresponds to a map $S^3\to S^3$ and there
are topologically distinct vacuum classes, since $\pi_3(S^3)=\bZ$.

In fact, it is possible to perform a gauge
transformation on the NCL (\ref{NCL_S}) which changes the asymptotic
behavior of the gauge fields, so that they can be considered to live on
$S^3=\bR^3\cup \{\infty\}$. Let $\omega(\mu, r, \theta, \phi)$ be an
$SU(2)$-valued
map which approaches $U(\mu, \theta, \phi)$ for $r\to \infty$ and $\id_2$
for $r\to 0$. The radial dependence of $\omega(\mu, r, \theta, \phi)$
implements a path which connects the
map $U(\mu, \theta, \phi)=\omega(\mu, \infty, \theta, \phi)$ to the constant map
$\id_2=\omega(\mu, 0, \theta, \phi)$. [Note that $U(\mu, \theta, \phi)$
for \emph{fixed $\mu$} is a map $S^2 \to S^3$ and therefore contractible.]

The crucial point, now, is that the map $\omega(0, r, \theta, \phi)$
is homotopically different from the map $\omega(\pi, r, \theta, \phi)$.
[Otherwise, the radial dependence of $\omega(\mu, r, \theta, \phi)$
would yield a contraction of $\omega(\mu, \infty, \theta, \phi)$,
considered as a $\mu$-dependent map $S^3\to S^3$, which is impossible.]
Both maps $\omega(0, r, \theta, \phi)$ and $\omega(\pi, r, \theta, \phi)$
can also be viewed as maps $S^3 \to S^3$, since
$\omega(\mu, \infty, \theta, \phi)=U(\mu, \theta, \phi)=\id_2$
for $\mu=0$ and $\mu=\pi$.  The conclusion is then that the corresponding
vacuum configurations $A_m(x^1,x^2,x^3)$ at $\mu=0$ and $\mu=\pi$
belong to different homotopy classes.
This result will be discussed further in Section \ref{sec:ChiralAnomaly}.

\subsection{Sphaleron $\mathrm{S}^*$}\label{sec:sphaleronSstar}

The three-space of our $SU(2)$ Yang--Mills--Higgs theory (\ref{actionYMH})
is again compactified by adding the ``sphere at
infinity.'' This time, however, we do not consider one-parameter families
(loops)  of static finite-energy configurations but two-parameter families
(spheres).
At spatial infinity, these families are characterized by the map
\begin{equation} \label{UNCSgeneral}
U: S^2 \times S^2 \to SU(2)\,, \qquad (\mu, \nu, \theta, \phi) \mapsto
U(\mu,\nu,\theta,\phi)\,,
\end{equation}
where ($\mu$, $\nu$) are the parameters of the sphere of configurations
and ($\theta$, $\phi$) are the polar and azimuthal angles of the spherical
coordinates in three-space. The parameters $\mu$ and  $\nu$ run from $-\pi/2$
to $+\pi/2$ and the boundary
of the ($\mu$,$\nu$)-square at $|\mu|=\pi/2$ or $|\nu|=\pi/2$ is
mapped to the same element of $SU(2)$.

Next, restrict the class of
mappings $U$ by requiring that $U(\mu, \nu, 0, \phi)$ is
independent of $(\mu$, $\nu$, $\phi)$ and $U( -\pi/2,
-\pi/2, \theta, \phi)$ independent of $(\theta$, $\phi)$. Then $U$ is
effectively a mapping from $S^4$ to $S^3$, which has a nontrivial
homotopy structure, $\pi_4(S^3) \eq \bZ_2$.
The general idea, now, is to construct a
noncontractible sphere, to determine the maximal energy configuration on
that sphere and to continuously deform the sphere so that its
maximal energy is minimized \cite{K85}.

The construction of the required nontrivial map $S^4 \to S^3$ is done in two
steps. First, a nontrivial map $S^3\to S^2$ is found and, second,
an operation is performed to increase the dimension of both spheres.

The relevant map $S^3 \to S^2$ is given by the well-known
\emph{Hopf fibration} \cite{Nak90}, which can be explained as follows.
Consider the three-sphere  $S^3$ to be a subset of $\bC^2$, namely
$\{(z_1,z_2)\,|\; |z_1|^2 + |z_2|^2 =1 \}$.
Each $\bC$-line through the origin in $\bC^2$ then
intersects with this three-sphere in a great circle $S^1$.
These great circles $S^1$ form a pairwise disjoint covering of $S^3$.
Two points of $S^3$ are defined to be equivalent ($\simeq$), if they lie
on the same great circle $S^1$. The corresponding projection,
\begin{equation}
S^3 \to S^3/\simeq \,,
\end{equation}
is the desired Hopf map, since the topological space $S^3/\simeq\:$
is homeomorphic to $S^2$.

The topological equivalence of $S^3/\simeq\:$ and $S^2$
can be shown by considering the $\bC$-lines which label
the great circles $S^1$ discussed in the previous paragraph. All
but one of these $\bC$-lines can be parametrized by complex numbers
$c\in \bC$. Specifically, the coordinates of such a line read
\begin{equation}
\left(z_1\, , \, z_2\right) =
(w\, , \, c \: w)\,,
\qquad   \mathrm{for}\;\;w \in \bC\,.
\end{equation}
In addition, there is the single $\bC$-line given by
\begin{equation}
\left(z_1\,,\, z_2\right) = (0\,,\, w)\,,
\qquad \mathrm{for}\;\;w  \in \bC\,.
\end{equation}
Hence, the total parameter space of $S^3/\simeq$ is given by the
one-point-compactified plane, i.e., the Riemann sphere $S^2$.

The  \emph{suspension} of a sphere $S^n$ is essentially the same as
the smash product $S^1 \wedge S^n$. It can be used to increase the
dimension of the spheres appearing in the above discussion. The resulting
suspension of the Hopf map corresponds
to a nontrivial element of the homotopy group  $\pi_4(S^3)=\bZ_2$.

In a particular parametrization, the required map (\ref{UNCSgeneral})
takes the form \cite{K85}
\begin{eqnarray} \label{USstar}
U(\mu,\nu,\hat x) & = & \bigl( \,\sin\mu + \i \cos\mu \, \,
  \exp [\,+\i (\nu+\pi/2)\, \tau_3\,]\,\,  \hat x \cdot \vec \tau\,\,
  \exp [\,-\i (\nu+\pi/2)\, \tau_3\,] \, \bigr) \nonumber\\
&&\times \bigl( \,\sin\mu - \i \cos\mu \,\, \hat x \cdot\vec\tau \,\bigr) \,,
\end{eqnarray}
where $\mu$ and  $\nu$ range over $[-\pi/2, \pi/2]$ and describe
a two-sphere, as does the unit three-vector
$\hat x \equiv (\sin\theta \cos\phi, \sin\theta \sin\phi,\cos\theta)$.
The map $U(\mu,\nu,\hat x)$ is effectively defined on the smash product
$S^2 \wedge S^2$, since $U$ is independent of $\hat x$ for $\mu=\pm \pi/2$
or $\nu=\pm \pi/2$ and independent of $\mu$ and $\nu$ for $\hat x =(0,0,1)$.
(Note that the suspended Hopf map also plays a role in the physics of
Skyrme solitons \cite{WZ83}.)

With the map (\ref{USstar}) in hand, it is possible to construct a
noncontractible sphere (NCS) of
static \YMH~configurations and to obtain the corresponding
nontrivial classical solution, the sphaleron $\mathrm{S}^*$,
just as for the NCL and the sphaleron $\mathrm{S}$ of the previous
subsection. The construction of $\mathrm{S}^*$ is,
however, rather subtle. Here, we only describe the four basic
steps and refer the reader to Ref.~\cite{K93} for more information.

First, we observe that the map (\ref{USstar}) singles out the $x^3$ axis,
which suggests the use of the cylindrical coordinates $\rho$, $\phi$, and $z$,
defined by $(x^1,x^2,x^3)=(\rho\cos\phi, \rho\sin\phi, z)$.
Then, it is not difficult to construct a NCS of
static $SU(2)$ \YMH~configurations, whose behavior at infinity is
governed by the $SU(2)$ matrix (\ref{USstar}). Specifically, the NCS
configurations can be written in terms of
six axial functions $f_i(\rho,z)$ and $h_j(\rho,z)$, for $i=0,1,2,3$ and
$j=1,2$, with appropriate \bcs~(for example, $f_0(\rho,z) \to 1$ and
$h_1(\rho,z) \to 1$ as $\rho^2 +z^2 \to \infty$). The
$SU(2)$ gauge and Higgs field configuration of the NCS are by construction
axially symmetric.

\begin{figure}[t]
\begin{center}
\includegraphics[width=3.5cm]{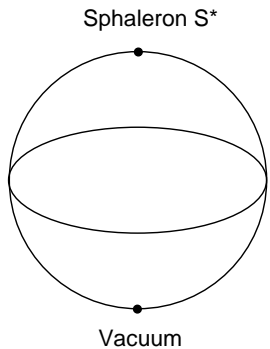}\\[8mm]
\end{center}
\caption{Sphaleron $\mathrm{S}^*$ on top of a noncontractible sphere in
  configuration space.}
\label{fig:sphaleronSstar}
\end{figure}

Second, the configuration at $\mu=\nu=0$ is also invariant under parity
reflection and gives the maximum energy of the NCS. Moreover, it
can be verified that this $\mu=\nu=0$ configuration, in terms of the six
axial functions $f_i(\rho,z)$ and $h_j(\rho,z)$,
provides a \emph{self-consistent Ansatz} for
the $SU(2)$ \YMH~field equations.
Concretely, this means that the \emph{Ansatz} reduces the field
equations to precisely six  partial differential equations (PDEs) for the
six functions $f_i(\rho,z)$ and $h_j(\rho,z)$, with appropriate
\bcs~which trace back in part to the finite-energy condition.
(This result agrees with the so-called
principle of symmetric criticality \cite{P79}, which states that,
under certain conditions, it suffices to consider variations that
respect the symmetry of the \emph{Ansatz}.) The solution of these PDEs
then determines the field configurations of the sphaleron
$\mathrm{S}^*$. See Fig.~\ref{fig:sphaleronSstar} for a sketch of
configuration space.

Third, the reduced field equations for the sphaleron $\mathrm{S}^*$ can be
solved numerically. For approximately vanishing quartic Higgs coupling constant
($\lambda/g^2 = 1/800$), the numerical solution of the six PDEs with the
correct \bcs~gives the following value for the energy:
\begin{equation}  \label{ESstar}
E_{\mathrm{S}^*} \approx 1.91 \times E_\mathrm{S}\,,
\end{equation}
where $ E_\mathrm{S}$ denotes the corresponding
energy of the sphaleron $\mathrm{S}$
[cf. Eq.~(\ref{ES}) above]. In fact, the sphaleron $\mathrm{S}^*$
is found to have the structure of a di-atomic molecule,
binding together a sphaleron  $\mathrm{S}$ and an ``anti-sphaleron''
$\bar{\mathrm{S}}$. See Ref.~\cite{K93} for a plot of the energy density
and further discussion.

Fourth, the construction of $\mathrm{S}^*$ via the NCS
can be extended to the full $SU(2) \times U(1)$ theory
of the \ewSM~by the introduction of one more axial function,
$f_4(\rho,z)$, with trivial boundary conditions at infinity.
But for nonvanishing weak mixing angle $\theta_w$, there are only
preliminary numerical results for the sphaleron $\mathrm{S}^*$  and it
would be worthwhile to obtain accurate
results over the full range of values of $\lambda/g^2$ and $\theta_w$.

\subsection{$Z$-string}

\label{sec:sphaleronZ}

Now consider static field configurations of the $SU(2)$
Yang--Mills--Higgs theory (\ref{actionYMH})
which are independent of one spatial
coordinate, the $z$-coordinate, and have vanishing gauge potential in
that direction, $A_z=0$.
In order to have finite total energy, the $z$-direction has to be compact
and three-space is taken to be $\bR^2 \times S^1$ instead of $\bR^3$.
Also, choose cylindrical polar coordinates ($\rho$, $\phi$, $z$) and work in
the polar gauge for which $A_\rho=0$.

Since the energy density in a
plane with fixed $z$ has to be finite, the remaining gauge field
component $A_\phi$ reduces to a pure-gauge configuration asymptotically,
\begin{equation}
g\, A_\phi \to - \, (\partial_\phi\, \omega)\,\,
\omega^{-1}\,, \quad \mathrm{as}\;\;\rho \to \infty\,,
\end{equation}
for a map $\omega: S^1 \to SU(2)$. Basically, this means that the plane
$\bR^2$ is compactified by adding the ``circle at infinity.''

It is possible to construct a noncontractible sphere of these field
configurations by restricting the corresponding maps
\begin{equation} \label{UZNCSgeneral}
U: S^2 \times S^1 \to SU(2)\,, \qquad
 (\mu, \nu, \phi) \mapsto U(\mu, \nu, \phi)\,,
\end{equation}
in such a way that they are effectively defined on the smash product $S^2
\wedge S^1 \sim S^3$. Specifically, the sphere is parametrized by
$\mu$ and $\nu$ which take values in $[-\pi/2, +\pi/2]$. The rim of the
($\mu$,$\nu$)-square is identified and corresponds to a single point on
$S^2$. The map $U$ is restricted to be independent of $\phi$ if $(\mu,
\nu)$ lies on this rim and independent of ($\mu$,$\nu$) if $\phi\eq 0$.

An appropriate expression for the map (\ref{UZNCSgeneral}) is
given by \cite{KO94}
\begin{subequations}
\label{UZNCS}
\begin{align}
U(\mu, \nu, \phi) & \eq V(\mu, \nu, 0)^{-1} \, V(\mu,
\nu, \phi)\,, \label{ZNCSptilde}\\
V(\mu, \nu, \phi) &\eq  z^1\, (-\i \tau_1) + z^2\, (-\i \tau_2)
+ z^3\, (-\i \tau_3) + z^4\, \id_2\,,  \\
\vect{z^1\\z^2\\z^3\\z^4} &\eq \vectleft{
\sin\mu \\ \cos\mu \sin\nu \\ \cos\mu \cos\nu \sin\phi\\ \cos\mu
\cos\nu \cos\phi} \,.
\end{align}
\end{subequations}
Note that the factor $V(\mu, \nu, 0)^{-1}$ in (\ref{ZNCSptilde}) serves
a dual purpose. First, it assures that the rim of the
($\mu$,$\nu$)-square is mapped to a single element, since $V(\mu,\nu,\phi)$
is independent of $\phi$ on this boundary. Second, it makes
$U$ independent of $\mu$ and $\nu$ if $\phi\eq 0$.

For the two-dimensional $SU(2)$ gauge and Higgs fields of the
noncontractible sphere (NCS), we make the \emph{Ansatz} \cite{KO94}
\begin{subequations}
\label{ZNCSansatz}
\begin{align}
g\, A(\mu,\nu,\rho,\phi) &= - f(\rho)\:{\rm d} U(\mu,\nu,\phi)\, U(\mu,\nu,\phi)^{-1}\,, \\
 \Phi(\mu,\nu,\rho,\phi) &= \eta \, h(\rho) \, U(\mu,\nu,\phi) \vect{0\\1}\,,
 \label{ZansPhi}
\end{align}
\end{subequations}
with parameters $|\mu|, |\nu| \leq \pi/2$.
The polar functions $f$ and $h$ have the following boundary conditions:
\begin{subequations} \label{ZNCSbcs}
\begin{align}
\lim_{\rho \to 0} \, f(\rho)/\rho&=0\,, &
\lim_{\rho \to \infty} \, f(\rho)&=1\,, \\
\lim_{\rho \to 0} \, h(\rho)&=0\,, &
\lim_{\rho \to \infty} \, h(\rho)&=1\,.
\end{align}
\end{subequations}

But no point on this NCS corresponds to a vacuum configuration, since the
Higgs field in Eq.\ (\ref{ZansPhi}) vanishes at $\rho=0$ for all
values of $(\mu,\nu)$. Therefore, the point of
the NCS at the boundary of the ($\mu$,$\nu$)-square must be connected
to the vacuum
by an additional line segment. The corresponding \emph{Ansatz}\/ is simply
\begin{subequations}\label{ZNCSrope}
\begin{align}
g\,A(\mu,\nu,\rho,\phi) &=0 \,, \\
\Phi(\mu,\nu,\rho,\phi) &=
     \eta\,\Bigl( 1 - \sin[\mu\nu] + h(\rho)\, \sin[\mu\nu] \, \Bigr)  \,
     \vect{0\\1}\,,
\end{align}
\end{subequations}
for $[\mu\nu] \equiv \max\{|\mu|, |\nu| \} > \pi/2$
with the parameter range of $\mu$ and $\nu$ extended to $[-\pi,+\pi]$.
The set of configurations (\ref{UZNCS})--(\ref{ZNCSrope}) is like a ``balloon''
which is tied to the ground by a rope.

\begin{figure}[t]
\begin{center}
\includegraphics[width=3.5cm]{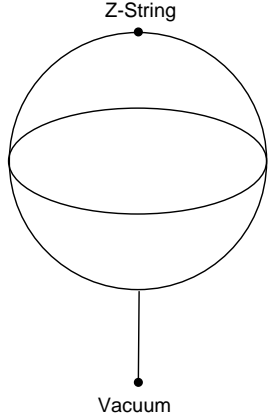}
\end{center}
\caption{$Z$-string on top of a noncontractible
         sphere (balloon) in configuration space.}
\label{fig:sphaleronZ}
\end{figure}

The energy of the NCS has a global maximum at $\mu=\nu=0$. By minimizing this
maximal energy with respect to the functions $f$ and $h$,
one finds the coupled differential equations
\begin{subequations}
\begin{align}
\rho\, f'' - f'         &= \half\, g^2 \eta^2\,\rho\, h^2\, (f-1) \,, \\
\rho^2\, h'' +\rho\, h' &= h\, (1-f)^2 + 2\lambda\eta^2\,\rho^2 h\, (h^2-1)\,,
\end{align}
\end{subequations}
where the prime indicates a derivative with respect to $\rho$.
The same differential equations, with  boundary conditions (\ref{ZNCSbcs}),
hold for the so-called $Z$-string \cite{NO73,N77,JPV93},
which excites the $Z\,$ boson and Higgs fields of the electroweak Standard
Model.

The $Z$-string is thus the sphaleron on the NCS given by
Eqs.~(\ref{UZNCS})--(\ref{ZNCSrope}); see Fig.~\ref{fig:sphaleronZ}.
This particular sphere (balloon) in configuration space will be discussed further
in Section~\ref{sec:ZstringAnomaly}. Note, finally, that the
configurations of the NCS can also be embedded in the full
$SU(2) \times U(1)$ gauge theory of the \ewSM;
see Ref.~\cite{KO94} for details and numerical results.

\section{Spectral flow}
\label{sec:sflow}
The classical field configurations of the previous section may
serve as background fields for massless Dirac fermions, whose
 left-handed components form an isodoublet under the $SU(2)$ gauge
 group and whose right-handed
components are gauge singlets. The Dirac equation for the spinor
$\Psi (x)$ reads in this case
\begin{equation}
\left(\i \Dsl - k\,  (\Phi_M^\dagger P_L + \Phi_M P_R)\right) \Psi =0\,,
\label{Diraceq}
\end{equation}
with the Yukawa coupling constant $k$, the Feynman slash notation
$\Dsl\equiv \gamma^\mu D_\mu$, and the covariant derivative
\begin{equation}
D_\mu \Psi (x) \equiv \left[\,\partial_\mu + g A_\mu (x) P_L \,\right] \Psi (x)\,,
\label{covderivchiral}
\end{equation}
which shows that only the left-handed fermions interact with the $SU(2)$ gauge field.

The left- and right-handed projectors are, as usual, defined by
$P_L \equiv \half (\id - \gamma_5)$
and  $P_R \equiv \half (\id +\gamma_5)$.
With the Minkowski metric of Section \ref{sec:YMH}, the Dirac matrices
obey the following Clifford algebra and Hermiticity conditions:
\begin{equation}
\gamma^\mu\gamma^\nu + \gamma^\nu\gamma^\mu = 2\, g^{\mu\nu}\,, \;\;\;
\gamma^{\mu\,\dagger} = \gamma^0\gamma^\mu\gamma^0        \,, \;\;\;
\gamma_5 \equiv \i \,\gamma^0\gamma^1\gamma^2\gamma^3 =\gamma_5^\dagger\,.
\label{Diracmatrices}
\end{equation}
(For the Euclidean metric, all Dirac matrices are chosen Hermitian,
$\gamma^{\mu\,\dagger} = \gamma^\mu$.)
The spacetime manifold considered in this article is flat and there
is no need to use the vierbeins (tetrads) explicitly.

The matrix $\Phi_M$ in (\ref{Diraceq}) contains the two Higgs field
components $ \Phi_1$ and $ \Phi_2$,
\begin{equation}
\Phi_M = \twomat{\Phi_2^* & \Phi_1 \\ -\Phi_1^* & \Phi_2}\,,
\end{equation}
so that
\begin{equation}
\Phi_M \cdot \vect{0\\1} = \vect{\Phi_1\\\Phi_2} = \Phi \,. \label{PhiMdef}
\end{equation}
For the Higgs vacuum with $A_\mu(x)=0$ and $\Phi_M (x)=\mathrm{diag}(\eta,\eta)$,
the effective fermion mass is given by $m=k\,\eta$.

The model considered may serve as the starting point for a consistently
renormalized quantum field theory with gauge group $SU(2)\times U(1)$
if we include three colors of left-handed
quark isodoublets for each left-handed lepton isodoublet,
so that the perturbative gauge
anomalies \cite{A69,BJ69,AB69,B69} cancel between the quarks
and the leptons \cite{BIM72,GJ72}.
(A similar cancellation occurs for the nonperturbative $SU(2)$
anomaly \cite{W82} to be discussed in Section \ref{sec:WittenAnomaly}.)
But, for our purpose, it suffices to consider a \emph{single}
isodoublet of left-handed
fermions, since the fermion isodoublets of the full theory behave identically.

The time-dependent solutions of the Dirac
equation (\ref{Diraceq}) are, however, not our main interest. Rather, we are
interested in the eigenvalues $E$ of the corresponding Dirac Hamiltonian,
\begin{equation} \label{HDirac}
H = - \i \gamma^0 \gamma^m D_m
    +k \,\gamma^0 \left(\,\Phi_M^\dagger P_L + \Phi_M P_R\,\right) \,,
\end{equation}
where use has been made of the fact that $A_0=0$ for our gauge
field configurations and
the covariant derivative $D_m$, for $m=1,2,3$, has already been
given in Eq.~(\ref{covderivchiral}). The eigenvalues $E$ are real,
since the Dirac Hamiltonian $H$ is Hermitian.

Now consider periodic one-parameter families (loops) of static background fields.
The spectral flow invariant \cite{APS75}
is then defined as the number of eigenvalues
that cross zero from below minus the number of eigenvalues that cross zero from
above as the loop parameter varies over its range (in a prescribed direction).
See, e.g.,  Ref.~\cite{A85} for an elementary introduction to the concept
of spectral flow.

Even if the spectral flow invariant vanishes, there may still be a
nontrivial rearrangement (permutation) of the energy levels. We speak
about ``spectral flow'' also in this case. (Mathematicians would perhaps
say that there is no spectral flow if the spectral flow invariant is zero.)
In addition, we will look for ``spectral flow'' in two-parameter families of
background fields (which may be characterized by a different topological invariant).

\subsection{Spectral flow for the sphaleron $\mathrm{S}$}
\label{sec:sflowS}

\begin{figure}[t]
\begin{center}
\includegraphics[height=4.5cm]{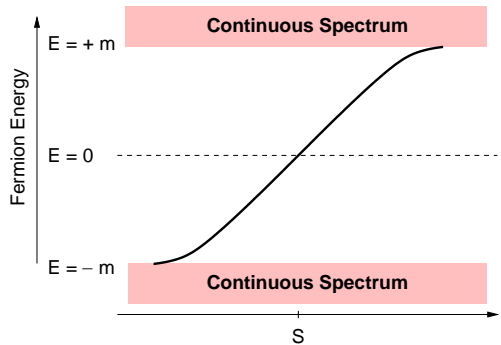}
\end{center}
\caption{Spectral flow for a path over the sphaleron barrier
  (cf. Fig.\ \ref{fig:sphaleronS}).}
\label{fig:sflowS}
\end{figure}

Consider the noncontractible loop (NCL) used in Section
\ref{sec:sphaleronS} to construct the sphaleron $\mathrm{S}$, with
parameter $\mu$ running from $0$ to $\pi$. At the beginning of the loop ($\mu=0$)
and at the end ($\mu=\pi$), the static background field (\ref{NCL_S})
is the same vacuum
configuration and the spectrum of the Dirac Hamiltonian (\ref{HDirac})
is purely continuous with a mass gap $2m$ according to
the Higgs mechanism ($m\propto\eta$).
For the sphaleron $\mathrm{S}$ at $\mu=\pi/2$, on the other hand,
it has been shown \cite{N75,JR76,KM84aps,R88,KB93} that the Dirac Hamiltonian
$H$ has a single normalizable eigenfunction with eigenvalue zero.

The overall picture, starting from $\mu=0$, is that a negative
eigenvalue $E(\mu)$ of the Dirac Hamiltonian $H$ rises above
the negative continuous spectrum, crosses zero when the background
fields pass the sphaleron barrier ($\mu=\pi/2$), and finally reaches the positive
continuous spectrum for $\mu=\pi$. See Fig.~\ref{fig:sflowS} for a sketch
and  Ref.~\cite{KB93} for numerical results.

The nonvanishing spectral flow over the NCL is guaranteed by the
Atiyah--Singer index theorem \cite{AS68}, which relates the analytic
index of the
four-dimensional Dirac operator (the loop parameter $\mu$ playing the
role of an imaginary time) to the  topological charge associated with the
NCL. Further details will be given in Section~\ref{sec:ChiralAnomaly}.
Here, we only remark that the NCL gauge field (\ref{NCL_S}), defined in
Minkowski space, has essentially the same topology as the BPST instanton
solution of Euclidean Yang--Mills theory \cite{BPST75,BC78}.

\subsection{Spectral flow for the sphaleron $\mathrm{S}^*$}
\label{sec:sflowSstar}

For the fermion behavior over the noncontractible sphere (NCS) through
$\mathrm{S}^*$, we need to resort to more abstract reasoning, since no
complete numerical or analytic solution has been obtained up till now.

First, consider massless Dirac fermions with
\emph{equal} gauge couplings for the right- and left-handed components.
It has then been shown  that there
exist two fermion zero-modes of the four-dimensional Euclidean Dirac operator
$\i \Dsl_4$, one of each chirality, if the fermions are placed in the background
of the constrained instanton $\mathrm{I}^*$ \cite{K93npb,KW96,K98npb}.
(Note that a particular time slice through $\mathrm{I}^*$ corresponds to the
three-dimensional configuration of the $\mathrm{S}^*$ sphaleron.
For practical purposes, one may consider $\mathrm{I}^*$ as a bound state
of a BPST  instanton $\mathrm{I}$ and an anti-instanton
$\bar{\mathrm{I}}$,
just as the sphaleron $\mathrm{S}^*$ may be viewed as a composite of a
sphaleron $\mathrm{S}$ and an anti-sphaleron $\bar{\mathrm{S}}$;
see Eq.~(\ref{ESstar}) and the lines below.)

Now the instanton $\mathrm{I}^*$, which depends on four Euclidean spacetime
coordinates, can be viewed as a path in configuration space which
passes over the $\mathrm{S}^*$ barrier. (In other words, this path
is homotopic to a particular closed loop on
the $\mathrm{S}^*$-NCS modulo gauge transformations;
cf. Fig.~\ref{fig:sphaleronSstar}.)
The two zero-modes of $\i \Dsl_4$ in the $\mathrm{I}^*$ background,
being time-dependent solutions of the
Dirac equation (with imaginary time), can be calculated in the
adiabatic approximation, where the state at time $t$ is an
eigenstate of the Dirac Hamiltonian with energy $E(t)$. The corresponding
``phase factor'' is given by
\begin{equation}
\exp \left( - \int_0^t \! {\rm d}t'\, E(t')  \right) \,.
\end{equation}
 From the normalizability of the zero-mode, it follows that
$E(t)$ is positive for $t \to +\infty$ and negative for $t \to -\infty$.

\begin{figure}[t]
\begin{center}
\includegraphics[height=4.5cm]{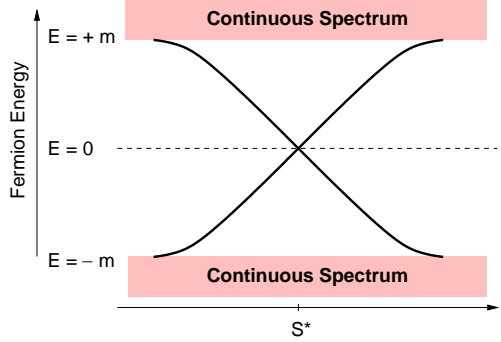}
\end{center}
\caption{Spectral flow for a path through the sphaleron $\mathrm{S}^*$
         (cf. Fig.\ \ref{fig:sphaleronSstar}).}
\label{fig:sflowSstar}
\end{figure}

With left- and right-handed chiralities, there are then two energy
levels $E^{(1,2)}(t)$
crossing zero from below (these energy levels may, of course, be degenerate).
In addition, there are two eigenvalues $E^{(3,4)}(t)$ which
cross zero from above, so that the total spectral flow invariant is zero
(note that the loop through $\mathrm{S}^*$ over the NCS is contractible).
For these last two eigenvalues, there are no
zero-modes of $\i \Dsl_4$ because the corresponding four-dimensional wave
functions are not normalizable. Thus we have two pairs of levels which
cross at zero energy, one left-handed pair and one right-handed pair.
Returning to the Dirac Hamiltonian (\ref{HDirac}) with only
left-handed fermions interacting with the $SU(2)$ gauge fields,
we have the spectral flow of the eigenvalues $E^{(1)}(t)$ and $E^{(3)}(t)$
as shown in Fig.~\ref{fig:sflowSstar}.

Recently, numerical results \cite{B03} have been obtained for the
eigenvalues of the $\i \Dsl_4$ operator along a particular path over the
$\mathrm{I}^*$ barrier.  It would be interesting to use similar methods to
calculate the spectral flow related to $\mathrm{S}^*$ and also to consider
fermion representations other  than isodoublet.

\subsection{Spectral flow for the $Z$-string}\label{sec:sflowZ}

Finally, we turn to the fermion behavior over the
noncontractible sphere (NCS) with the $Z$-string at the top \cite{KR97}.
First, we choose a path on the NCS (\ref{UZNCS})--(\ref{ZNCSrope}), which
starts in the vacuum, passes over the $Z$-string and ends up in the vacuum.

\begin{figure*}[t]
\begin{center}
\includegraphics[height=4.5cm]{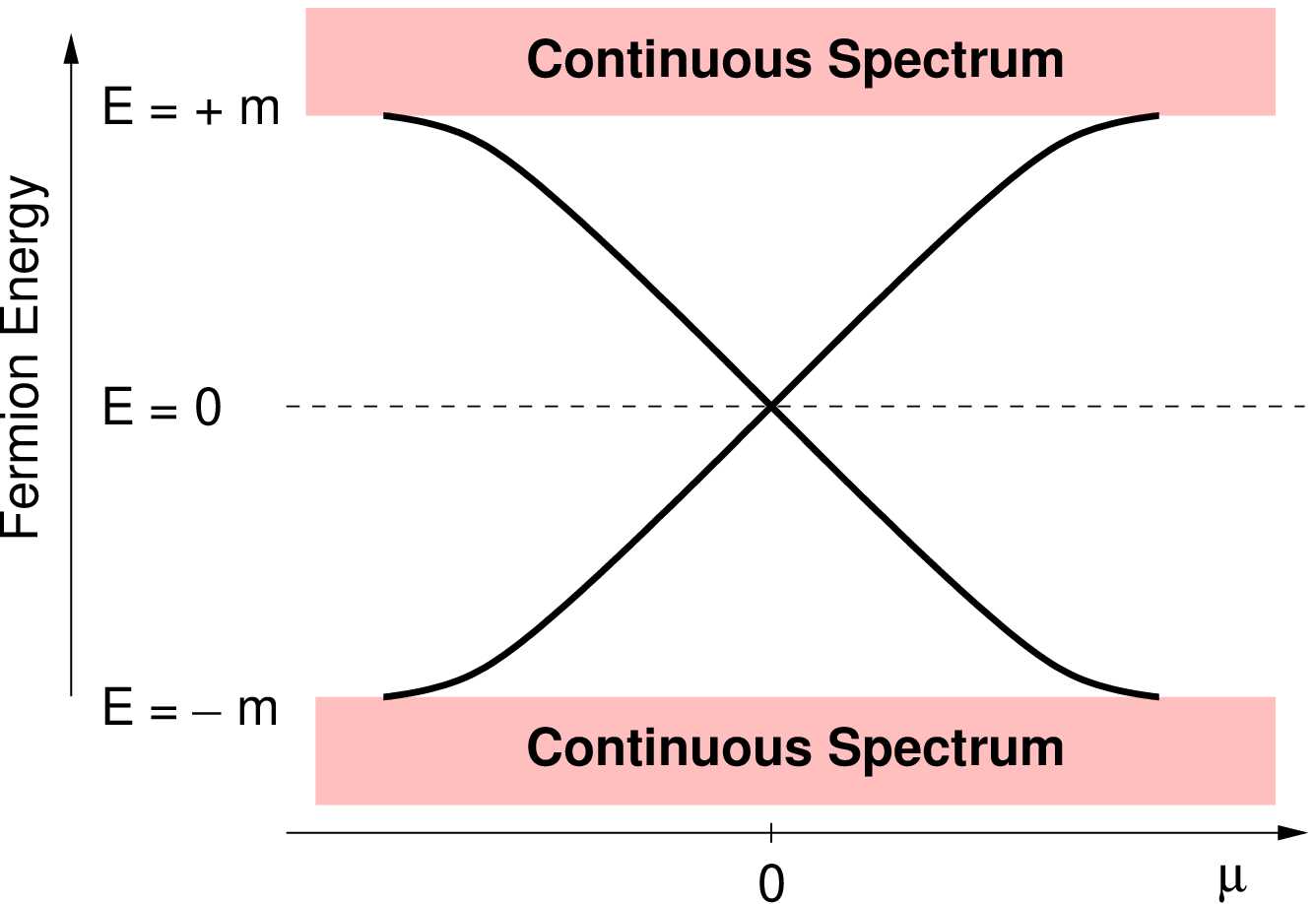}
\hspace{0.4cm}
\raisebox{5.7cm}{\includegraphics[width=6cm,angle=270]{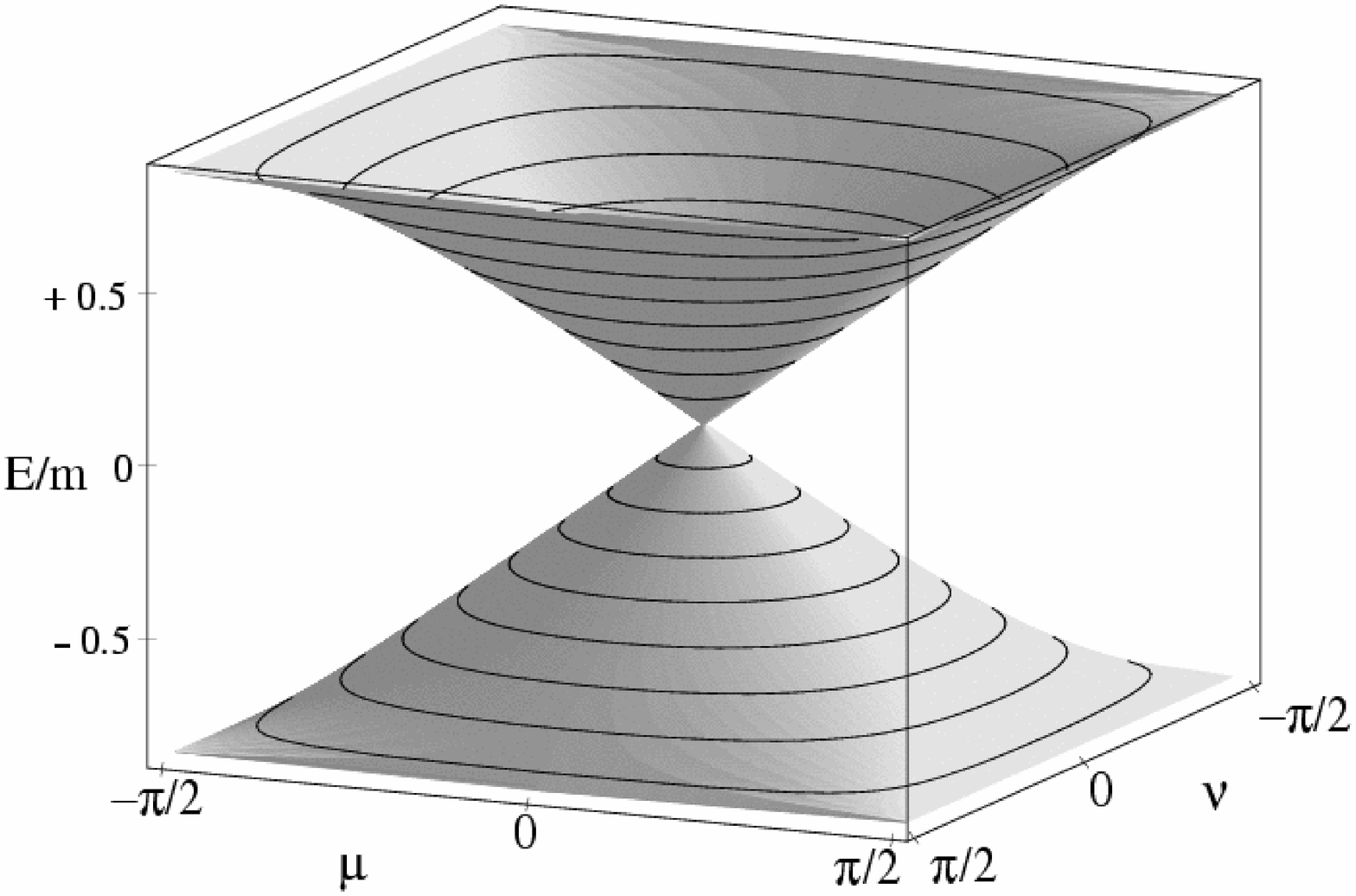}}
\end{center}
\caption{Spectral flow over the noncontractible sphere (NCS) through the
  $Z$-string (cf. Fig.~\ref{fig:sphaleronZ}). In the picture on the left,
  the NCS parameter $\nu$ is held fixed at the value $0$ and $\mu$ is varied.
  In the picture on the right, the Dirac eigenvalues are shown for
  the NCS patch $|\mu|, |\nu| \le \pi/2$.}
\label{fig:sflowZ}
\end{figure*}

To be concrete, we put $\nu=0$ in (\ref{UZNCS}) and let $\mu$ run from
$-\pi/2$ to $+\pi/2$.
Such a loop is contractible and there is no net spectral flow to be
expected (just as for the loop through $\mathrm{S}^*$ considered in the
previous subsection). What happens instead is that one negative eigenvalue
$E^{(1)}(\mu,0)$ is raised from the negative continuous spectrum and
one positive eigenvalue $E^{(2)}(\mu,0)$
is lowered from the positive continuous spectrum.
Both eigenvalues meet at energy zero when the background
fields pass the $Z$-string configuration ($\mu=\nu=0$), cross and
reach the opposite region of continuous eigenvalues
(see the picture on the left in Fig.~\ref{fig:sflowZ}). The fermion
zero-modes of the $Z$-string have been studied in Refs.~\cite{JR81,EP94}.

We can also consider the behavior of the Dirac eigenvalues over
the whole two-parameter family (\ref{ZNCSansatz}). Plotted over the
($\mu$,$\nu$)-square, the eigenvalues $E(\mu,\nu)$ form a double cone
meeting at $\mu=\nu=0$ (see the picture on the right in Fig.~\ref{fig:sflowZ}).

\section{Anomalies}\label{sec:anomalies}

In this section, we review the relation between the sphalerons presented
in Section \ref{sec:sphalerons} and
so-called anomalies.
The connection between sphalerons and anomalies is precisely
the spectral flow discussed in Section \ref{sec:sflow}.

\subsection{Chiral anomaly and fermion number violation}
\label{sec:ChiralAnomaly}

The chiral  $U(1)$ anomaly, which turns out to be related to the sphaleron
$\mathrm{S}$, eliminates a rigid $U(1)$
symmetry of the classical action,  viz.  chiral invariance.
This anomaly can be found in theories with \emph{massless} fermions,
for which there is a classical Ward identity
\begin{equation}
\sum_f \left(\, \fdq{\Gcl}{\Psi_f} \; \delta \Psi_f -
              \delta \bar \Psi_f \; \fdq{\Gcl}{\bar\Psi_f}\,\right) =
\partial^\mu j_\mu^5\,, \label{chirWIclass}
\end{equation}
where $\Gcl$ is the classical action and $\Psi_f(x)$ denotes a Dirac
fermion field, with $f$ labeling the different flavors (fermion species).
The rigid chiral transformation of the fermion fields is given by
\begin{equation}
\delta \Psi_f (x)= \i \gamma_5\, \Psi_f(x)\,, \qquad \delta \bar \Psi_f(x) =
 \bar \Psi_f(x)\, \i\gamma_5\,.
\end{equation}
Since the left-hand side of (\ref{chirWIclass}) vanishes for
solutions of the classical equations of motion, the current
$j_\mu^5(x) \equiv \sum_f \bar \Psi_f(x) \,\gamma_5 \gamma_\mu \Psi_f(x)$
is conserved classically. This implies that
the chiral charge $Q_5\equiv \int {\rm d}^3x \, j_0^5(x)$
does not change with time ($t \equiv x^0$),
\begin{equation} \label{Q5conservation}
\left. \frac{\mathrm{d}Q_5}{\mathrm{d}t} \right|_\mathrm{\,classical}=0\,.
\end{equation}

\begin{figure}[t]
\begin{center}
\includegraphics[height=4cm]{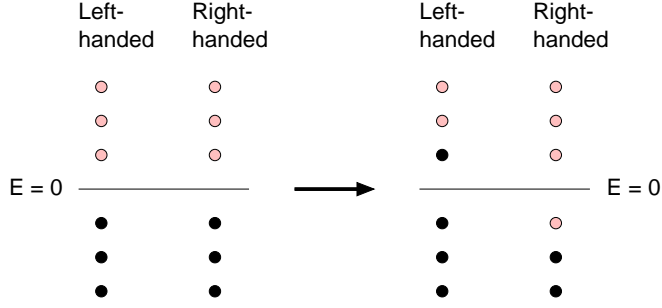}
\end{center}
\caption{\label{fig:chiranom}Opposite spectral flow for
  left- and right-handed fermions, which leads to the creation of two units of
  chiral charge. Filled states are drawn black, empty states gray.}
\end{figure}

Now suppose that the $SU(2)$ gauge field couples equally to left- and
right-handed fermions in the fundamental representation
(as is the case for the $SU(3)$ gauge field
which is believed to be responsible for quark confinement in the \SM).
Then the spectral flow for a path over the
sphaleron $\mathrm{S}$ with unit winding number is as shown in
Fig.\ \ref{fig:chiranom}: for each isodoublet of fermions a left-handed
state crosses zero from below and a right-handed one crosses zero from
above. (Essentially the same type of spectral flow has
been found \cite{M85} in the Schwinger model, i.e.,
two-dimensional quantum electrodynamics with a massless Dirac fermion.)
In the Dirac-sea picture of the second-quantized vacuum, this
means that a pair of fermions is created from an initial vacuum state,
namely one chiral fermion and one chiral antiparticle
corresponding to a hole in the Dirac sea of antichiral negative-energy
states. Hence, the total chiral charge $Q_5$ \emph{changes}
by two units per isodoublet, which contradicts the classical conservation
equation (\ref{Q5conservation}).

This result is supported by the Atiyah--Singer index theorem \cite{AS68}
for the four-dimensional chiral Dirac operator
(see, e.g., Refs.~\cite{JR77,F80,B00}). For $N$ isodoublets,
the relation between the change of chiral charge and the appropriate
characteristic of the background gauge field
is simply the integrated version of the
perturbative Ward identity for the chiral current
containing the Adler--Bell--Jackiw anomaly \cite{A69,BJ69},
\begin{equation}
\sum_{f=1}^{N} \left(\, \fdq{\Gamma}{\Psi_f} \; \delta \Psi_f  -
              \delta \bar \Psi_f   \;  \fdq{\Gamma}{\bar\Psi_f} \,\right)
= \left[\, \partial^\mu j_\mu^5 \,\right]\bullet \Gamma
\,+\, \frac{g^2\,N}{8\,\pi^2} \,\,
\left[ \tr\,\,  \tilde F_{\mu\nu}\, F^{\mu\nu} \right]\bullet \Gamma
\,, \label{chirWIquant}
\end{equation}
where $\Gamma$ is now the fully quantized vertex functional and the
bullet denotes an operator insertion.

The anomalous term in Eq.~(\ref{chirWIquant}) includes the Pontryagin density
\begin{equation}
q(x) \equiv - \,\frac{g^2}{16\, \pi^2}\, \tr\,\, \tilde F_{\mu\nu}(x)\, F^{\mu\nu}(x)\,,
\end{equation}
with
$\tilde F_{\mu\nu} \equiv (1/2) \, \epsilon_{\mu\nu\rho\sigma}F^{\rho\sigma}$.
The integral of this density over the spacetime manifold $M$
is a  \emph{topological invariant} called the Pontryagin index,
\begin{equation}
Q \equiv \int_{M} {\rm d}^4x\;  q(x)  \,.
\end{equation}
 For compact spacetime manifolds $M$, the Pontryagin index is an integer number
and is also called the winding number or topological ``charge''
(hence, the notation $Q$).

\begin{figure}[t]
\begin{center}
\includegraphics[height=4.5cm]{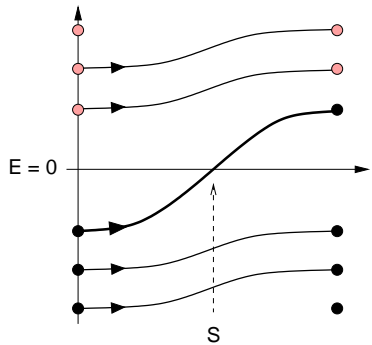}
\end{center}
\caption{\label{fig:fnumanomaly}Spectral flow in the electroweak
  Standard Model, where only left-handed
  fermions interact with the $SU(2)$ gauge fields.
  $\mathrm{S}\,$ denotes the sphaleron, which has a single fermion zero-mode.
  The spectral flow leads to a change of fermion number between initial
  and final states (see text).}
\end{figure}

Next, turn to the simplified version of the \ewSM, as described in
Sections \ref{sec:YMH} and \ref{sec:sflow}. Here, the
fermion fields are fundamentally massless, even though they behave
as massive particles in the Higgs vacuum.
More importantly, the gauge field now couples only to the
left-handed parts of the fermion fields; cf.\ Eqs.\ (\ref{Diraceq}) and
(\ref{covderivchiral}).
Hence, the spectral flow for a single fermion flavor
is made up of only one state which crosses zero
from below. This implies that \emph{fermion number conservation} is violated
\cite{tH76prl,tH76prd}.
See Fig.~\ref{fig:fnumanomaly} and compare with Fig.~\ref{fig:chiranom}.
(It is, of course, important to define carefully what is meant by
``the fermion number'' of a given state \cite{C80,GH95,Kh95}; see also the
discussion in the last three paragraphs of this subsection.)

The map $U$ given in Section \ref{sec:sphaleronS} essentially
provides a map $S^3 \to S^3$,
characterized by the topological charge $|Q| = 1$.  The above
considerations can be generalized to other (integer) values of $Q$ and to
a model with $N_{\rm fam}$ families of quarks and leptons.
The sum of baryon number $B$ and lepton number $L$
is then found to be nonconserved \cite{tH76prl},
\begin{equation}
\Delta(B-L) = 0\,,\quad \Delta(B+L) = 2\, N_{\rm fam} \, Q \,, \label{DBL}
\end{equation}
where $\Delta B$ denotes the change of baryon number between
initial and final states and similarly for $\Delta L$.

As explained at the end of Section \ref{sec:sphaleronS}, the NCL can
be transformed into a path connecting two topologically distinct vacua
in one-point-compactified three-space. The general form of such a
vacuum is given by a static pure-gauge configuration,
\begin{equation}
g\, A = -  {\rm d}\chi \, \chi^{-1}\,, \qquad
\Phi  = \eta\, \chi \vect{0\\1}\,,
\end{equation}
for a map $\chi: \bR^3 \to SU(2)$ which approaches $\id_2$
at spatial infinity. The homotopy class to which $\chi$ belongs is
characterized by the integer \emph{Chern--Simons number}
\begin{equation}
N_\mathrm{CS}[\, \chi\,]= -
   \,\frac{1}{24\,\pi^2} \int {\rm d}^3x\: \epsilon^{klm} \,\, \tr\left\{
  (\partial_k \chi\,\chi^{-1}) \,
  (\partial_l \chi\,\chi^{-1}) \,
  (\partial_m \chi\,\chi^{-1}) \right\} \,.
\end{equation}
The topological charge $Q$ of the map $\omega(\mu,r,\theta,\phi)$,
as discussed in the last two paragraphs of Section \ref{sec:sphaleronS},
is then the difference of the
Chern--Simons numbers of the vacua at the start and end of the
associated path,
\begin{equation}
Q = \Delta N_\mathrm{CS}\equiv
    N_\mathrm{CS}[\, \omega(\pi,r,\theta,\phi)\, ]-
    N_\mathrm{CS}[\, \omega(0,r,\theta,\phi)\, ]\,.
\label{QNCS}
\end{equation}
Of course, it is also possible to map the $\mu$-interval $[0,\pi]$
on the time interval $[-\infty,+\infty]$.

\begin{figure}[t]
\begin{center}
\includegraphics[height=3.3cm]{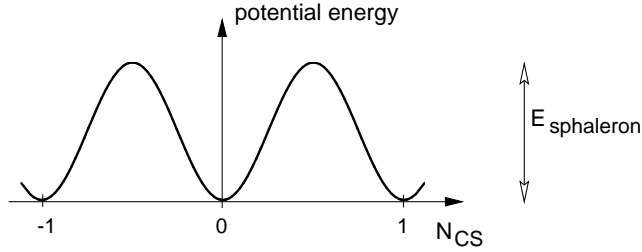}
\end{center}
\caption{\label{fig:barrier}Potential energy over a slice of
  configuration space, parametrized by
  the   Chern--Simons number  $N_\mathrm{CS} $. The height of the energy barrier
  between   topologically different vacua ($N_\mathrm{CS} = n \in \bZ$) is
  set by the sphaleron S, which appears in different gauge copies
  ($N_\mathrm{CS} = 1/2+ n$, for $n \in \bZ$).
  This figure essentially ``unwraps'' the loop of Fig.~\ref{fig:sphaleronS}.}
\end{figure}

The sphaleron now corresponds to an energy barrier between
these vacua, as sketched in Fig.\ \ref{fig:barrier}.
The transition between different vacua can, for example, take place by
tunneling \emph{through} the sphaleron barrier \cite{tH76prl,tH76prd}
or by passing \emph{over} the barrier due to a
thermal fluctuation of the fields \cite{KM84,DS78}.
Especially the latter mechanism is expected to
contribute significantly to fermion-number-violating processes in the early
universe (see, e.g., Refs.\ \cite{McL94,RS96}).

The rate of fermion-number-violating processes at relatively low energies
($E\ll E_\mathrm{S} \approx 10\;\mathrm{TeV}$)
can be calculated from the Euclidean path integral \cite{tH76prd,R90}.
But for a reliable discussion of these processes at high energies
($E \gtrsim E_\mathrm{S}$) it is advisable to remain in
Minkowski spacetime. The problem, then, is that the
compactification of four-space which we used as the starting point of
our topological considerations is not really physically sensible for
Minkowski spacetime. The topological charge $Q$, in particular,  need not be
an integer quantity in Minkowski spacetime. The crucial point here is
the role of \emph{energy conservation} for background fields that
solve the equations of motion; see, e.g., Ref.~\cite{FKS93}.
The general question of which type of gauge field leads to nontrivial spectral
flow remains unanswered for the moment.

There exists, however,  a result for \emph{strongly dissipative}
$SU(2)$ gauge fields \cite{C80,GH95,Kh95}. In this case, the spectral flow
is given by the difference in winding numbers of the asymptotic vacuum
configurations for $t \to \pm \infty$. Roughly speaking, this coincides with the
previous result in Euclidean spacetime, namely Eq.~(\ref{QNCS})
inserted into Eq.~(\ref{DBL}).

For the case of spherically symmetric fields, there is also a result
for \emph{generic} (i.e., nondissipative) gauge fields,
\begin{equation}
\Delta(B-L) = 0\,, \quad \Delta(B+L) = 2 \, N_{\rm fam} \,
\left(
\Delta N_\mathrm{winding}
+ \Delta N_{\rm twist}\right)\,
\Bigr|_\mathrm{spher.\; symm.}  \,.
\end{equation}
The change of $B+L$ is now determined by two integers.
The first, $\Delta N_\mathrm{winding}$, again corresponds to
Eq.~(\ref{QNCS}). But the second, $\Delta N_{\rm twist}$, is an entirely
new characteristic of spherically symmetric $SU(2)$  gauge fields,
which is related to the asymptotic behavior of the solutions of a
(nonlinear) Riccati equation embedded in the (linear) zero-energy
Dirac equation \cite{KL01}. The integer $\Delta N_{\rm twist}$ is zero for
strongly dissipative $SU(2)$ gauge fields.
See Ref.~\cite{K02blois} for further discussion of the issues involved.

\subsection{Witten's global gauge anomaly}\label{sec:WittenAnomaly}

The global $SU(2)$ gauge anomaly, which turns out to be  related to the sphaleron
$\mathrm{S}^*$, differs from the case discussed in the previous subsection in that
not just a symmetry of the theory is eliminated but the theory itself.

As mentioned in Section \ref{sec:sflowSstar}, the crossing of
energy levels for paths over the $\mathrm{S}^*$ barrier is related
to the existence of two normalizable zero-modes of the four-dimensional
Euclidean Dirac operator $\i \Dsl_4$, one of each chirality.
The noncontractible sphere of three-dimensional configurations can
also be viewed as  a noncontractible loop of four-dimensional
configurations. Furthermore, as explained at the end of Section
\ref{sec:sphaleronS}, it is possible to pass  from  a \emph{loop}
of gauge field configurations in the radial gauge to a \emph{path}
of gauge field configurations without the radial gauge condition.
The resulting path has topologically inequivalent vacua at the start
and at the end.

Now consider the change of eigenvalues of $\i \Dsl_4$
along such a path. Since for one ``point'' of the path (i.e., the
$\mathrm{I}^*$-configuration) there are known zero-modes
\cite{K98npb}, it is to be expected that some level crossing is
occurring also here.

\begin{figure*}[t]
\begin{center}
\includegraphics[height=4.5cm]{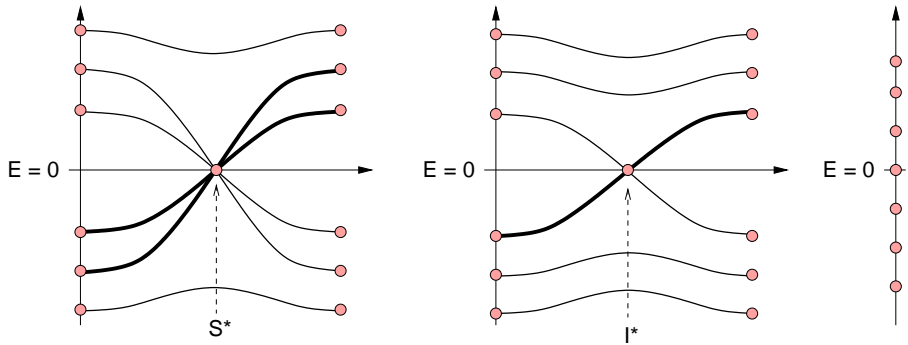}
\end{center}
\caption{Left: sketch of the eigenvalues of the Dirac Hamiltonian for
  a particular path over the $\mathrm{S}^*$ noncontractible sphere
  (the theory considered has both left- and right-handed
  fermions interacting with the $SU(2)$ gauge fields,  which is not the case
  in Fig.~\ref{fig:sflowSstar}).
  Middle: eigenvalues of $\i \Dsl_4$ for the corresponding noncontractible
  loop through the constrained instanton $\mathrm{I}^*$.
  Right:  eigenvalues of $\i \Dsl_5$ for the corresponding
  five-dimensional configuration.   Each path with an energy level crossing zero
  from below (thick line) corresponds to a normalizable fermion zero-mode one
  dimension higher. The sphaleron $\mathrm{S}^*$ has four fermion zero-modes,
  the constrained instanton $\mathrm{I}^*$  two,
  and the five-dimensional configuration one.}
\label{fig:WittenAnomaly}
\end{figure*}

That this is indeed the case has been shown in
Ref.~\cite{W82} by increasing the dimension once more. The
one-parameter family of four-dimensional Dirac operators can also be
considered as a single five-dimensional one. (In other words, the whole NCS
serves as a single background configuration.) It then follows from the so-called
mod--2 Atiyah--Singer index theorem \cite{AS71} that the corresponding
five-dimensional Dirac operator has a normalizable zero-mode.
For $\i \Dsl_4$, this implies that an eigenvalue is crossing zero from negative
to positive values as the path is traversed. Simultaneously, there is a
second eigenvalue which passes from positive to negative values.
This discussion is summarized in Fig.\ \ref{fig:WittenAnomaly},
which also gives the corresponding spectral flow in three
dimensions. (The mod--2 index theorem guarantees only an odd number of
zero-modes for the five-dimensional configuration,
but for simplicity we have assumed there is just one. See Ref.~\cite{B03}
for numerical results and further discussion.)

Witten also argued that the spectral flow of
$\i \Dsl_4$ leads to a global gauge anomaly \cite{W82}.
In the Euclidean path integral of
$SU(2)$ Yang--Mills theory with a single isodoublet of Weyl fermions,
there effectively appears a square root of the Dirac determinant,
\begin{equation}
  \int {\D}A_\mu \;\; \sqrt{\det \i \Dsl_4}\;\; \exp \left(\,
  \frac{1}{2} \int \dx \tr\, F_{\mu\nu} F^{\mu\nu} \,\right)\,,
\label{pathint}
\end{equation}
if one recalls that two Weyl fermions of opposite chiralities make a single
Dirac fermion.

The Dirac determinant in Eq.~(\ref{pathint})
depends on the background gauge fields $A_\mu (x)$
and its square root can be defined as the product of the positive eigenvalues
[\,starting from a given gauge field configuration, say $A_\mu (x)=0\,$].
The above considerations then show that for a particular
\emph{continuous} variation of the gauge
fields we end up with gauge fields, which are related to the
starting configuration by a large gauge transformation and which have a
$\sqrt{\det \i \Dsl_4}$ of opposite sign (one positive eigenvalue
having become negative; cf. the middle picture of
Fig.~\ref{fig:WittenAnomaly}). In the path integral,
one has to integrate over all gauge fields (taking out the infinite
factor due to gauge invariance afterwards). Hence, for every contribution
$\sqrt{\det \i \Dsl_4}$ there is also a contribution
$-\sqrt{\det \i \Dsl_4}$ arising from the gauge-transformed background fields.
This implies that the path integral (\ref{pathint}) vanishes.

More precisely, the path integral over the gauge fields is not well defined,
because there is no satisfactory way to restrict the integration over
the gauge fields so that a single Weyl isodoublet gives a continuous
gauge-invariant contribution. This, then, is the Witten anomaly, which
can also be proven without the mod--2
Atiyah--Singer index theorem but with the perturbative Bardeen
anomaly instead \cite{EN84,K91plb}.

\subsection{$Z$-string global gauge anomaly}
\label{sec:ZstringAnomaly}

\begin{figure*}[t]
\begin{center}
\includegraphics[height=3.5cm]{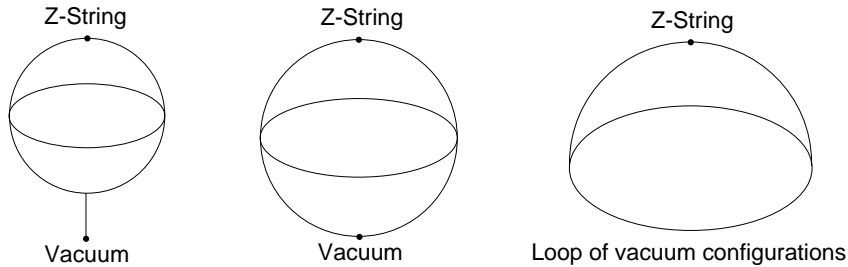}
\end{center}
\caption{$Z$-balloon, $Z$-sphere and $Z$-disc.}
\label{fig:Zdisc}
\end{figure*}

Just as for the Witten anomaly and the $\mathrm{S}^*$  sphaleron
of the previous subsection, there exists a
global gauge anomaly related to the $Z$-string sphaleron \cite{KR97}.
In order to explain this anomaly, we need a modified noncontractible sphere (NCS),
obtained by continuous deformation of
the balloon as given in Section \ref{sec:sphaleronZ}. This modification
has the advantage of being a real sphere, that is, without
degenerate points. The modified NCS still has one point corresponding to the
vacuum and one point corresponding to the $Z$-string
(see the picture in the middle of Fig.\ \ref{fig:Zdisc}).

For the modified NCS, the $z$-independent $SU(2)$ gauge field is
in the polar gauge $A_\rho=0$. But like the case of the sphalerons
$\mathrm{S}$ and $\mathrm{S}^*$, it is possible to relax the polar gauge
condition and to demand instead that the vacuum reached for $\rho \to
\infty$ is the trivial one. Then one ends up with a disc of
configurations  with a loop of pure-gauge configurations on the
boundary (see the picture on the right of Fig.\ \ref{fig:Zdisc}).
Considering the compactified radial coordinate $\rho$
to be a polar angle $\theta$, the fields are effectively defined on
a sphere $S^2$. The loop of vacuum configurations on this two-sphere, restricted
to the smash product $S^1 \wedge S^2$, corresponds to a nontrivial element of
the homotopy group $\pi_3(S^3)$.

We keep this in mind for later and turn to the eigenvalues of the
four-dimensional Euclidean Dirac operator $\i \Dsl_4$, where the time-dependent
background fields are taken to be paths over the $Z$-disc, with the start and
end point (not necessarily the same) lying on the rim of
vacuum configurations. For any such path passing through the
$Z$-string, we know from Section \ref{sec:sflowZ} that $\i \Dsl_4$ has a
single normalizable zero-mode corresponding to the eigenvalue of the
Dirac Hamiltonian which crosses zero from below.

\begin{figure}[t]
\begin{center}
\includegraphics[height=4cm]{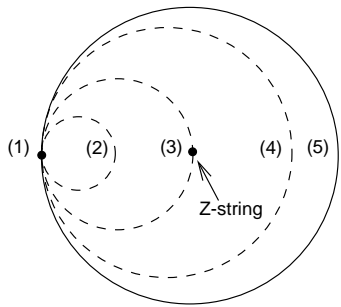} \qquad
\end{center}
\vspace{-0.25\baselineskip}
\caption{Five closed loops on the $Z$-disc (cf. Fig.~\ref{fig:Zdisc}).
  Loop (1) corresponds to a single point on the boundary,
  loop (3) passes through the $Z$-string, and loop (5) consists of the
  whole boundary of the disc.}
\label{fig:Zdiscloops}
\end{figure}

Now consider a particular
family of operators $\i \Dsl_4$ corresponding to a family of paths over
the $Z$-disc, which starts from a constant path corresponding to a point on
the rim of the disc,
passes through a path via the $Z$-string, and ends up in a pure
vacuum path formed by the boundary of the disc
(see Fig.~\ref{fig:Zdiscloops}, where the $Z$-disc of Fig.\ \ref{fig:Zdisc}
has been flattened). This family
of four-dimensional Dirac operators sweeps
over the whole $Z$-disc and we expect that there is spectral flow
corresponding to the winding
number of the underlying map $S^3 \to S^3$. In our case, this means
that a single eigenvalue of $\i \Dsl_4$ crosses zero. The zero crossing
can be expected to occur for the path labeled $(3)$ in  Fig.~\ref{fig:Zdiscloops}.

Like the case of the Witten anomaly in Section~\ref{sec:WittenAnomaly},
this prevents us from defining the square
root of the fermion determinant as a continuous gauge-invariant function of the
bosonic background fields. Note that our path of four-dimensional configurations
begins in a time-independent, topologically trivial vacuum and ends up in a
gauge-transformed, time-dependent and topologically nontrivial  one.

It is particularly interesting to see how this global gauge anomaly
manifests itself in the space of time-independent fermion
states. Since the Dirac Hamiltonian is a \emph{real} Hermitian operator,
we may choose real energy-eigenstates. Concretely, look at the
one-dimensional subspace spanned by the eigenstate
which crosses zero from below in the picture on the left in
Fig.\ \ref{fig:sflowZ}. For the background fields, we use an arbitrary
loop on the $Z$-disc, which circumnavigates the $Z$-string exactly once and which
is parametrized by $\alpha \in [0, 2\pi]$. Then the
energy eigenstate defines a \emph{real} line bundle over $S^1$.

\begin{figure}[t]
\begin{center}
\includegraphics[height=5cm, angle=270, bb= 172 202 460 569]{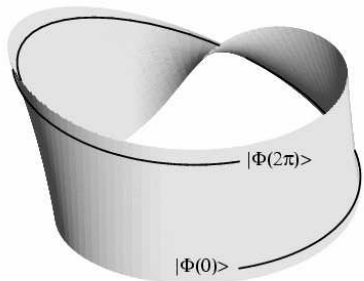}
\qquad
\end{center}
\caption{M\"{o}bius bundle structure of the gauge orbit, with loop parameter
  $\alpha \in [0, 2\pi]$. The line
  represents a normalized real eigenstate $|\Phi(\alpha)\rangle$ of the Dirac
  Hamiltonian, with $|\Phi(2\pi)\rangle=-|\Phi(0)\rangle$.}
\label{fig:moebius}
\end{figure}

It has been shown in Ref.~\cite{KR97} that this bundle is, in fact, the
M\"{o}bius bundle. A normalized eigenstate $|\Phi(0)\rangle$ transported
around the loop ends up as $|\Phi(2\pi)\rangle=-|\Phi(0)\rangle$; see
Fig.\ \ref{fig:moebius}. The phase factor found is determined by the
Berry phase for adiabatic transport \cite{B84}. The Berry phase
$\pi \pmod{2\pi}$ is of topological origin and does not change under
continuous deformation of the loop, as long as the loop of configurations
does not touch the fermion degeneracy ``point'' corresponding to the
$Z$-string. This observation also shows that the boundary of the $Z$-disc
(Fig.~\ref{fig:Zdisc}) is a noncontractible loop of vacuum configurations, since
the real Berry phase factor $-1$ on it cannot be continuously changed to $+1$.

The variation of the
eigenstate along the rim of the $Z$-disc defines a projective action of
the gauge group on the fermionic matter.
There is then a global gauge anomaly,
because it is impossible to define a real, continuous, and proper
(i.e., nonprojective) representation of the
local gauge group on the fermion states. Since the vacuum of quantum
field theory is  the Dirac sea with all negative-energy
eigenstates filled, this also means that the second-quantized
vacuum state  acquires the Berry phase factor $-1$. See Section 6 of
Ref.~\cite{KR97} for further discussion. (We
take the opportunity to correct a slip of the pen.
In the last sentence of Footnote 6 in Ref. \cite{KR97},
the words ``and vice versa'' must be deleted.)

A similar interpretation of the Witten anomaly in terms of a Berry phase
has been given in Ref.~\cite{NA85}.
There is, however, a crucial difference between the $Z$-string global
gauge anomaly and the Witten anomaly. For the $Z$-string anomaly,
namely, there \emph{does} exist a local counterterm in the action
which restores gauge invariance, but at the price of violating Lorentz
and CPT invariance \cite{K98npb2}. More generally, if gauge invariance is
enforced, there appears a new anomaly, the so-called CPT anomaly
(see Refs.\ \cite{K00npb,KS02} for the main result
and Ref.~\cite{K01wigner} for a review).

\section{Conclusion} \label{sec:conclusion}

The space of finite-energy three-dimensional field configurations of
$SU(2)$ Yang--Mills--Higgs theory (in short, configuration space) has nontrivial
topology \cite{T82,M83}, which leads to the existence of a new type of
classical solutions, the so-called sphalerons.
Sphalerons are unstable static finite-energy solutions of the classical
field equations, whereas solitons are stable solutions.

In Section \ref{sec:sphalerons}, we have explained the
topology behind the $\mathrm{S}$, $\mathrm{S}^*$, and $Z$-string sphalerons
\cite{KM84,K93,KO94} of the $SU(2)$ Yang--Mills--Higgs theory (\ref{actionYMH}).
Precisely this theory appears in the electroweak Standard Model of
elementary particle interactions \cite{EWSM}. Knowledge of these classical
solutions may, therefore, be of great importance to physics.

Adding chiral fermions to the $SU(2)$ Yang--Mills--Higgs theory,
the nontrivial topology of
configuration space makes itself felt by the occurrence of spectral
flow \cite{APS75}, as discussed in Section \ref{sec:sflow}.
In turn, the general phenomenon of spectral flow
is related to the possible existence of anomalies
which invalidate certain properties of the classical theory, as discussed in
Section \ref{sec:anomalies}.

The spectral flow over a noncontractible loop through the
sphaleron $\mathrm{S}$ is related to the chiral $U(1)$ anomaly
\cite{A69,BJ69},
which corresponds to a breakdown of baryon and lepton number
conservation in the \ewSM~\cite{tH76prl}. The spectral
flow through the $\mathrm{S}^*$ and $Z$-string sphalerons does \emph{not}
lead to a global $SU(2)$ gauge anomaly \cite{W82,KR97},
because the electroweak Standard Model has
an \emph{even} number of chiral isodoublets. Still, there is nontrivial spectral
flow (more precisely, spectral rearrangement)
over configuration space, but its physical implications
remain to be clarified (cf.\ Refs.\ \cite{K98npb,K00npb}).
Indeed, we need a better understanding of the role of
configuration space topology in concrete physical problems, such as
the behavior of elementary particle fields at high energies or
temperatures.


\begin{thebibliography}{99}

\bibitem{Sk60}
T.H.R.~Skyrme, ``A non-linear field theory,''
Proc.\ R.\ Soc.\ London\ Ser.\ A {\bf   260} (1960) 127;
\\
T.H.R.~Skyrme,
``A unified field theory of mesons and baryons,''
Nucl.\ Phys.\ {\bf 31} (1962) 556.


\bibitem{tHP74}
G.~'t Hooft,
``Magnetic monopoles in unified gauge theories,''
Nucl.\ Phys.\ B {\bf 79} (1974) 276;
\\
A.M.~Polyakov,
``Particle spectrum in quantum field theory,''
JETP Lett.\  {\bf 20} (1974) 194
[Pisma Zh.\ Eksp.\ Teor.\ Fiz.\  {\bf 20} (1974) 430].

\bibitem{T82}
C.H.~Taubes,
``The existence of a non-minimal solution to the $SU(2)$
Yang--Mills--Higgs equations on $\bR^3$,''
Commun.\ Math.\ Phys.\  {\bf   86} (1982) 257; {\bf   86} (1982) 299;
\\
C.H.~Taubes,
``Min-max theory for the Yang--Mills--Higgs equations,''
Commun.\ Math.\ Phys.\  {\bf 97} (1985) 473.

\bibitem{M83}
N.S.~Manton,
``Topology in the Weinberg--Salam theory,''
Phys.\ Rev.\ D {\bf 28} (1983) 2019.


\bibitem{KM84}
F.R.~Klinkhamer and N.S.~Manton,
``A saddle point solution in the Weinberg--Salam theory,''
Phys.\ Rev.\ D {\bf 30} (1984) 2212.


\bibitem{K93}
F.R.~Klinkhamer,
``Construction of a new electroweak sphaleron,''
Nucl.\ Phys.\ B {\bf 410} (1993) 343
[arXiv:hep-ph/9306295].

\bibitem{KO94}
F.R.~Klinkhamer and P.~Olesen,
``A new perspective on electroweak strings,''
Nucl.\ Phys.\ B {\bf 422} (1994) 227
[arXiv:hep-ph/9402207].

\bibitem{EWSM}
S.L.~Glashow,
``Partial symmetries of weak interactions,''
Nucl.\ Phys.\  {\bf 22} (1961) 579;
\\
S.~Weinberg,
``A model of leptons,''
Phys.\ Rev.\ Lett.\  {\bf 19} (1967) 1264;
\\
A.~Salam,
``Weak and electromagnetic interactions,''
in: \emph{Elementary Particle Theory}, edited by N.~Svartholm
(Almqvist, Stockholm, 1968), p.~367;
\\
S.L.~Glashow, J.~Iliopoulos and L.~Maiani,
``Weak interactions with lepton--hadron symmetry,''
Phys.\ Rev.\ D {\bf 2} (1970) 1285.

\bibitem{McL94}
L.D.~McLerran,
``B+L nonconservation as a semiclassical process,''
Acta Phys.\ Polon.\ B {\bf 25} (1994) 309
[arXiv:hep-ph/9311239].


\bibitem{RS96}
V.A.~Rubakov and M.E.~Shaposhnikov,
``Electroweak baryon number non-con\-ser\-va\-tion in the early universe and in
high-energy collisions,''
Usp.\ Fiz.\ Nauk {\bf 166} (1996) 493 [Phys.\ Usp.\  {\bf 39} (1996) 461]
[arXiv:hep-ph/9603208].


\bibitem{Nak90}
M.~Nakahara, \emph{Geometry, Topology and Physics},
Institute of Physics Publishing, Bristol, 1990.

\bibitem{Mil63}
J.~Milnor, \emph{Morse Theory},
Princeton University Press, Princeton, 1963.

\bibitem{DHN74}
R.~Dashen, B.~Hasslacher and A.~Neveu,
``Nonperturbative methods and extended hadron models in field theory.
  III. Four-dimensional nonabelian models,''
Phys.\ Rev.\ {\bf D 10} (1974) 4138.


\bibitem{K90plb}
F.R.~Klinkhamer,
``Sphalerons, deformed sphalerons and configuration space,''
Phys. Lett. {\bf 236 B} (1990) 187.

\bibitem{KKB92}
J.~Kunz, B.~Kleihaus and Y.Brihaye, ``Sphalerons at finite mixing angle,''
Phys.\ Rev.\ D {\bf 46} (1992) 3587.

\bibitem{HJ94}
M.~Hindmarsh and M.~James,
``The origin of the sphaleron dipole moment,''
Phys.\ Rev.\ D {\bf 49} (1994) 6109
[arXiv:hep-ph/9307205].


\bibitem{KB89}
J.~Kunz and Y.~Brihaye,
``New sphalerons in the Weinberg-Salam theory,''
Phys.\ Lett.\ B {\bf 216} (1989) 353.

\bibitem{Y89}
L.G.~Yaffe,
``Static solutions of $SU(2)$ Higgs theory,''
Phys.\ Rev.\ D {\bf 40} (1989) 3463.


\bibitem{K85}
F.R.~Klinkhamer,
``A new sphaleron in the Weinberg-Salam theory?,''
Z.\ Phys.\ C {\bf 29} (1985) 153.

\bibitem{WZ83}
F.~Wilczek and A.~Zee,
``Linking numbers, spin, and statistics of solitons,''
Phys.\ Rev.\ Lett.\  {\bf 51} (1983) 2250.


\bibitem{P79}
R.S.~Palais,
``The principle of symmetric criticality,''
Commun.\ Math.\ Phys.\ {\bf 69} (1979) 19.


\bibitem{NO73}
H.B.~Nielsen and P.~Olesen,
``Vortex-line models for dual strings,''
Nucl.\ Phys.\ B {\bf 61} (1973) 45.

\bibitem{N77}
Y.~Nambu,
``String-like configurations in the Weinberg--Salam theory,''
Nucl.\ Phys.\ B {\bf 130} (1977) 505.

\bibitem{JPV93}
M.~James, L.~Perivolaropoulos and T.~Vachaspati,
``Detailed stability analysis of electroweak strings,''
Nucl.\ Phys.\  B {\bf 395} (1993) 534
[arXiv:hep-ph/9212301].


\bibitem{A69}
S.L.~Adler,
``Axial vector vertex in spinor electrodynamics,''
Phys.\ Rev.\  {\bf 177} (1969) 2426.


\bibitem{BJ69}
J.S.~Bell and R.~Jackiw,
``A PCAC puzzle: $\pi^0 \to \gamma \gamma$ in the sigma model,''
Nuovo Cimento\ A {\bf 60} (1969) 47.

\bibitem{AB69}
S.L.~Adler and W.A.~Bardeen,
``Absence of higher order corrections in the anomalous axial vector
divergence equation,'' Phys.\ Rev.\  {\bf 182} (1969) 1517.

\bibitem{B69}
W.A.~Bardeen,
``Anomalous Ward identities in spinor field theories,''
Phys.\ Rev.\  {\bf 184} (1969) 1848.

\bibitem{BIM72}
C.~Bouchiat, J.~Iliopoulos and P.~Meyer,
``An anomaly free version of Weinberg's model,''
Phys.\ Lett.\ B {\bf 38} (1972) 519.


\bibitem{GJ72}
D.J.~Gross and R.~Jackiw,
``Effect of anomalies on quasirenormalizable theories,''
Phys.\ Rev.\ D {\bf 6} (1972) 477.

\bibitem{W82}
E.~Witten,
``An $SU(2$) anomaly,''
Phys.\ Lett.\ B {\bf 117} (1982) 324.


\bibitem{APS75}
M.~Atiyah, V.K.~Patodi and I.M.~Singer,
``Spectral   asymmetry and Riemannian geometry'',
Math.\ Proc.\ Camb.\ Phil.\
Soc.\ {\bf 77} (1975) 43; {\bf 78} (1975) 405; {\bf 79} (1976) 71.

\bibitem{A85}
M.~Atiyah,
``Topological aspects of anomalies,''  in:
\emph{Symposium on Anomalies, Geometry, Topology}, edited by W.A.~Bardeen
and A.R.~White, World Scientific, Singapore, 1985, p. 22.

\bibitem{N75}
C.R.~Nohl,
``Bound state solutions of the Dirac equation in extended hadron models,''
Phys.\ Rev.\ D {\bf 12} (1975) 1840.

\bibitem{JR76}
R.~Jackiw and C.~Rebbi,
``Solitons with fermion number 1/2,''
Phys.\ Rev.\ D {\bf 13} (1976) 3398.

\bibitem{KM84aps}
F.R.~Klinkhamer and N.S.~Manton,
``A saddle-point   solution in the Weinberg--Salam theory,''
in: \emph{Proceedings of     the 1984 Summer Study on the Design and Utilization
of the     Superconducting Super Collider}, edited by R.~Donaldson and
J.~Morfin (American Physical Society, New York, 1984), p.~805.


\bibitem{R88}
A.~Ringwald,
``Sphaleron and level crossing,''
Phys.\ Lett.\ B {\bf 213} (1988) 61.


\bibitem{KB93}
J.~Kunz and Y.~Brihaye,
``Fermions in the background of the sphaleron barrier,''
Phys.\ Lett.\ B {\bf 304} (1993) 141
[arXiv:hep-ph/9302313].


\bibitem{AS68}
M.~Atiyah and I.M.~Singer,
``The index of elliptic  operators I, III, IV'',
Ann.\ Math.\ {\bf 87} (1968) 484;  {\bf 87} (1968) 546; {\bf 93} (1971) 119.

\bibitem{BPST75}
A.A.~Belavin, A.M.~Polyakov, A.S.~Schwartz and Y.S.~Tyupkin,
``Pseudoparticle solutions of the Yang-Mills equations,'' Phys.\
Lett.\ B {\bf 59} (1975) 85.

\bibitem{BC78}
K.M.~Bitar and S.J.~Chang,
``Vacuum tunneling of gauge theory in Minkowski space,''
Phys.\ Rev.\ D {\bf 17} (1978) 486.


\bibitem{K93npb}
F.R.~Klinkhamer,
``Existence of a new instanton in constrained Yang--Mills--Higgs theory,''
Nucl.\ Phys.\ B {\bf 407} (1993) 88
[arXiv:hep-ph/9306208].

\bibitem{KW96}
F.R.~Klinkhamer and J.~Weller,
``Construction of a new constrained instanton in Yang--Mills--Higgs theory,''
Nucl.\ Phys.\ B {\bf 481} (1996) 403
[arXiv:hep-ph/9606481].


\bibitem{K98npb}
F.R.~Klinkhamer,
``Fermion zero-modes of a new constrained instanton in Yang--Mills--Higgs  theory,''
Nucl.\ Phys.\ B {\bf 517} (1998) 142
[arXiv:hep-th/9709194].


\bibitem{B03}
O.~B\"ar,
``On Witten's global anomaly for higher $SU(2)$ representations,''
Nucl.\ Phys.\ B {\bf 650} (2003) 522
[arXiv:hep-lat/0209098].


\bibitem{KR97}
F.R.~Klinkhamer and C.~Rupp,
``A global anomaly from the $Z$-string,''
Nucl.\ Phys.\ B {\bf 495} (1997) 172
[arXiv:hep-th/9702023].



\bibitem{JR81}
R.~Jackiw and P.~Rossi,
``Zero modes of the vortex--fermion system,''
Nucl.\ Phys.\ B {\bf 190} (1981) 681.


\bibitem{EP94}
M.A.~Earnshaw and W.B.~Perkins,
``Stability of an electroweak string with a fermion condensate,''
Phys.\ Lett.\ B {\bf 328} (1994) 337
[arXiv:hep-ph/9402218].

\bibitem{M85}
N.S.~Manton,
``The Schwinger model and its axial anomaly,''
Ann.\ Phys.\ (N.Y.) {\bf 159} (1985) 220.

\bibitem{JR77}
R.~Jackiw and C.~Rebbi,
``Spinor analysis of Yang--Mills theory,''
Phys.\ Rev.\ D {\bf 16} (1977) 1052.


\bibitem{F80}
K.~Fujikawa,
``Path integral for gauge theories with fermions,''
Phys.\ Rev.\ D {\bf 21} (1980) 2848;
Erratum\ {\bf 22} (1980) 1499.


\bibitem{B00}
R.A.~Bertlmann, \emph{Anomalies in Quantum Field Theory},
Oxford University Press, Oxford, 2000.

\bibitem{tH76prl}
G.~'t Hooft,
``Symmetry breaking through Bell--Jackiw anomalies,'' Phys.\ Rev.\
Lett.\ {\bf 37} (1976) 8.


\bibitem{tH76prd}
G.~'t Hooft,
``Computation of the quantum effects due to a four-dimensional pseudoparticle,''
Phys.\ Rev.\ D {\bf 14} (1976) 3432;
Erratum\ {\bf 18} (1978) 2199.

\bibitem{C80}
N.H.~Christ,
``Conservation law violation at high-energy by anomalies,''
Phys.\ Rev.\ D {\bf 21} (1980) 1591.


\bibitem{GH95}
T.M.~Gould and S.D.~Hsu,
``Anomalous violation of conservation laws in Minkowski space,''
Nucl.\ Phys.\ B {\bf 446} (1995) 35
[arXiv:hep-ph/9410407].

\bibitem{Kh95}
V.V.~Khoze,
``Fermion number violation in the background of a gauge field in Minkowski space,''
Nucl.\ Phys.\ B {\bf 445} (1995) 270
[arXiv:hep-ph/9502342].

\bibitem{DS78}
S.~Dimopoulos and L.~Susskind,
``Baryon number of the universe,''
Phys.\ Rev.\ {\bf D 18} (1978) 4500.


\bibitem{R90}
A.~Ringwald,
``High-energy breakdown of perturbation theory in the electroweak
instanton sector,'' Nucl.\ Phys.\ B {\bf 330} (1990) 1;
\\
O.~Espinosa,
``High-energy behavior of baryon and lepton number violating scattering
amplitudes and breakdown of unitarity in the Standard Model,''
Nucl.\ Phys.\ B {\bf 343} (1990) 310;
\\
S.Y.~Khlebnikov, V.A.~Rubakov and P.G.~Tinyakov,
``Instanton induced cross-sections below the Sphaleron,''
Nucl.\ Phys.\ B {\bf 350} (1991) 441.



\bibitem{FKS93}
E.~Farhi, V.V.~Khoze, and R.~Singleton,
``Minkowski space nonabelian classical solutions with noninteger winding number change,''
Phys.\ Rev.\ {\bf D 47} (1993) 5551.


\bibitem{KL01}
F.R.~Klinkhamer and Y.J.~Lee,
``Spectral flow of chiral fermions in nondissipative gauge field  backgrounds,''
Phys.\ Rev.\ D {\bf 64} (2001) 065024
[arXiv:hep-th/0104096].


\bibitem{K02blois}
F.R.~Klinkhamer,
``Electroweak baryon number violation,'' in: \emph{XIV-th
  Rencontre de Blois: Matter--Antimatter Asymmetry}\/,
edited by  L.~Iconomidou-Fayard  and J.~Tran Thanh Van, in press
[arXiv:hep-ph/0209227].


\bibitem{AS71}
M.F.~Atiyah and I.M.~Singer,
``The index of elliptic   operators V,''
Ann.\ Math.\ {\bf 93} (1971) 139.


\bibitem{EN84}
S.~Elitzur and V.P.~Nair,
``Nonperturbative anomalies in higher dimensions,''
Nucl.\ Phys.\ B {\bf 243} (1984) 205.


\bibitem{K91plb}
F.R.~Klinkhamer,
``Another look at the $SU(2)$ anomaly,''
Phys.\ Lett.\ B {\bf 256} (1991) 41.


\bibitem{B84}
M.V.~Berry,
``Quantal phase factors accompanying adiabatic changes,''
Proc.\ R.\ Soc.\ London\ Ser.\ A {\bf 392} (1984) 45.


\bibitem{NA85}
P.~Nelson and L.~Alvarez-Gaum\'e,
``Hamiltonian interpretation of anomalies,''
Commun.\ Math.\ Phys.\  {\bf 99} (1985) 103.


\bibitem{K98npb2}
F.R.~Klinkhamer,
``$Z$-string global gauge anomaly and Lorentz non-invariance,''
Nucl.\ Phys.\ B {\bf 535} (1998) 233
[arXiv:hep-th/9805095].

\bibitem{K00npb}
F.R.~Klinkhamer,
``A CPT anomaly,''
Nucl.\ Phys.\ B {\bf 578} (2000) 277
[arXiv:hep-th/9912169].

\bibitem{KS02}
F.R.~Klinkhamer and J.~Schimmel,
``CPT anomaly: a rigorous result in four dimensions,''
Nucl.\ Phys.\ B {\bf 639} (2002) 241
[arXiv:hep-th/0205038].

\bibitem{K01wigner}
F.R.~Klinkhamer,
``CPT violation: mechanism and phenomenology,'' in:
 \emph{Proceedings of the Seventh International Wigner
  Symposium}, edited by M.E.~Noz, in press
[arXiv:hep-th/0110135].


\end{thebibliography}
\end{document}